\documentclass[12pt]{article}
\usepackage{fullpage,amsmath,amssymb,titling,xcolor,graphicx,hyperref}
\numberwithin{equation}{section}
\title{Defect entanglement entropy for superconformal monodromy defects}
\author{}
\date{}

\newcommand{\dd}{\mathrm{d}}

\begin{document}

\begin{titlepage}
\vfill

\begin{flushright}
CCTP-2025-13\\
ITCP-2025-13
\end{flushright}

\begin{center}

\vskip .5in 
\noindent

{\Large\bf{Defect entanglement entropy for superconformal monodromy defects}}

\bigskip\medskip

Andrea Conti$^{a,b}$\footnote{{\tt contiandrea@uniovi.es}}, Yolanda Lozano$^a$\footnote{{\tt ylozano@uniovi.es}}, Filippos Rogdakis$^c$\footnote{{\tt filippos.rogdakis@stud.uni-heidelberg.de}} \\
and Christopher Rosen$^c$\footnote{{\tt rosen@physics.uoc.gr}}

\bigskip\medskip
{\small 

\textit{$^a$Department of Physics, University of Oviedo,
Avda. Calvo Sotelo s/n, 33007 Oviedo}}

\smallskip
{\small \textit{and}}

\smallskip
{\small 

\textit{Instituto Universitario de Ciencias y Tecnolog\'ias Espaciales de Asturias (ICTEA),\\
Calle de la Independencia 13, 33004 Oviedo, Spain}}

\bigskip
{\small 

\textit{$^b$Blackett Laboratory, Imperial College,
Prince Consort Rd., London, SW7 2AZ, U.K}}

\bigskip
{\small 

\textit{$^c$Crete Center for Theoretical Physics, Department of Physics, University of Crete,\\
71003 Heraklion, Greece}}

\vskip 1cm 

     	{\bf Abstract }
	\end{center}
	We compute the defect entanglement entropy for co-dimension two superconformal monodromy defects in well known maximally symmetric holographic theories of various dimension. In each case we explicitly relate the universal part of the defect entanglement entropy to field theory data characterising the defect conformal field theory. We provide evidence that, unlike in the bulk theories in which the defects reside, the universal part of the defect entanglement entropy does not necessarily decrease along a renormalisation group flow.
	\noindent

\noindent

\vfill
\eject

\end{titlepage}

\tableofcontents

\newpage

\section{Overview}

Much as superconformal field theories provide an important computational handle on the physics of interacting quantum field theories, conformal defects which preserve some amount of supersymmetry present an interesting opportunity to efficiently characterize the physics of quantum field theories deformed by operators which break (some part of) the ambient theory's spacetime symmetries.

Understanding such deformations is particularly interesting in the context of the renormalization group (RG). While Poincar\'e invariant theories deformed by spatially homogenous couplings to relevant operators are known to exhibit monotonicity along an RG flow in certain dimension dependent observables, a similarly robust characterization of the behaviour of defect theories at different length scales remains lacking.

In this work, we focus our attention on a particular class of codimension two defects in superconformal field theories (SCFTs) of dimension $d=3,4$ and 6 on Minkowski space. These defect conformal field theories (dCFTs) preserve an $SO(d-2,2)\times SO(2)\subset SO(d,2)$ subgroup of the ambient theory's conformal group, as well as at least two Poincar\'e supercharges. They are characterized by non-trivial monodromies for Abelian subgroups of the global symmetry enjoyed by the theory, optionally accompanied by a conical deficit/excess at the location of the defect. 

The monodromy parameters, which we generically refer to as $g\mu^I$, can be understood as flat background connections $\mathcal{V}^I$ which couple to the conserved currents $\mathcal{J}_I$ of the global symmetry. From this perspective, the monodromy defect is described by a superconformal deformation of the ambient SCFT of the form
\begin{equation}
S_\mathrm{dCFT} = S_\mathrm{SCFT} + \int \mathcal{V}^I \wedge \star\, \mathcal{J}_I \qquad \mathrm{with} \qquad \mathcal{V}^I = g\mu^I \dd z.
\end{equation}

Here $z$ is an angular coordinate with period $\Delta z= 2\pi$ normal to the defect's worldvolume. Specifically, translations along $z$ realize the $SO(2)$ factor in the preserved $SO(d-2,2)\times SO(2)$ symmetry of the defect theory. Upon application of Stokes theorem, integrating the background gauge field around the defect one finds that the monodromy defect can be thought of as localised magnetic flux piercing the origin of the plane normal to the defect.  

Our focus in this work will be on understanding some features of superconformal monodromy defects in strongly interacting SCFTs. In particular, we extend the investigations initiated in \cite{Gutperle:2022pgw,Capuozzo:2023fll,Arav:2024wyg,Arav:2024exg} to include a study of the entanglement entropy between a spherical entangling region surrounding the defect and its complement on a fixed time slice. In each example, we are able to express the defect contribution to this entanglement entropy entirely in terms of field theory defect data---in particular, in terms of the monodromy sources $g\mu^I$ and a conical deficit parameter $n$ characterising the presence (or absence) of a conical singularity at the location of the defect.

Subsequently, we use our results together with those of \cite{Capuozzo:2023fll,Arav:2024wyg,Arav:2024exg} to express the defect's contribution to the entanglement entropy in terms of the so-called ``defect conformal weight'' $h_D$ which appears in the expectation value of the stress tensor in the defect SCFT (Weyl transformed to an AdS$_{d-1}\times \mathbb{S}^1$ background)
\begin{equation}
\langle \mathcal{T}_{ab} \rangle\dd x^a \dd x^b = -\frac{h_D}{2\pi}\left[\dd s^2(AdS_{d-1})-(d-1)n^2 \dd z^2 \right],
\end{equation}
and a remainder. In turn, the remainder is shown to be proportional to the defect free energy $I_D$ in three dimensions, and the defect Weyl anomaly coefficient $b$, $\mathcal{A}$ in four and six dimensions respectively. In even dimensions, this provides an explicit realisation of the general expectation for the defect entanglement entropy for codimension two conformal defects given in \cite{Jensen:2018rxu} and generalised in \cite{Chalabi:2021jud}. In odd dimensions we are unaware of analogous expectations.

From this reparametrisation of the defect entanglement entropy, we position ourselves to comment on the behaviour of the defect entanglement entropy along an RG flow. Distinct from the class of flows discussed in e.g. \cite{Jensen:2015swa,Casini:2016fgb,Cuomo:2021rkm}, we motivate the existence of an RG flow between particular defect SCFTs driven by spatially homogeneous mass deformations {\it away} from the defect. By explicitly evaluating the defect entanglement entropy in the putative UV and IR  defect theories, we discover that (unlike its analogue in the ambient theory without defects) the universal part of the entanglement entropy does not in general decrease along such a flow.

As a by-product of our analysis, we are further led to revisit the holographic renormalisation relevant for observables of the $d=6$ monodromy defects. This enables us to generalise the results of \cite{Bianchi:2021snj} which relate to the dependence of the defect anomaly coefficient $\mathcal{A}$ on the monodromy sources to include also the dependence on the conical singularity parameter $n$. 

The plan for the rest of the paper is as follows: in section \ref{sec:eCfun}, we briefly review the broadstroke features of the solutions holographically dual to the superconfonformal monodromy defects of interest, as well as the holographic prescription for computing the spherical entanglement entropy for entangling regions centered on the codimension two defect. We then define the defect entanglement entropy and introduce a simple formula which can be used to compute it in these backgrounds.

In sections \ref{sec:3d} and \ref{sec:4d}, we apply our formula to superconformal monodromy defects of the ABJM theory and $\mathcal{N}=4$ Super Yang-Mills respectively. In each case, we discuss the behaviour of the defect entanglement entropy under the putative renormalization group flows between superconformal defect theories driven by spatially homogeneous mass deformations away from the defect. In section \ref{sec:6d} we reproduce known results for the defect entanglement entropy for monodromy defects in the $6d$ $\mathcal{N}=(2,0)$ theory. We then combine these results with various observables computed via holographic renormalization to extend these results to systems with conical singularities. 

We conclude in section \ref{sec:dis}, providing further context for our results and discussing their consequences. Appendix \ref{app:sols} catalogues the solutions studied in this work, while appendix \ref{app:cut} describes in detail the holographic regularization and renormalization prescriptions required to obtain finite field theory observables. In appendix \ref{app:RG}, we observe that while the defect contribution to the entanglement entropy does not necessarily decrease along the RG flows introduced in sections \ref{sec:3d} and \ref{sec:4d}, in these systems there is a scheme independent part of the sphere entanglement entropy evaluated in those dCFTs ``at large'' which does.

\section{Entropic central charges from holography}\label{sec:eCfun}

\subsection{General considerations}
An efficient means of studying strongly coupled dCFTs is to holographically encode their symmetries in the isometries of a bulk gravitational solution. We accomplish this by studying gravitational solutions of type IIB and $D=11$ supergravity which can be realised in the maximal gauged supergravities that descend from these theories upon dimensional reduction. 

The solutions fall within an ansatz with metric of the form
\begin{equation}\label{eq:gAnz}
\dd s^2 = e^{2V}\dd s^2\left(\mathrm{AdS}_{d-1} \right) + h^2 \dd z^2 + f^2 \dd y^2
\end{equation}
in which $z$ is a coordinate on $\mathbb{S}^1$ with period $\Delta z = 2\pi$, and $V,h,$ and $f$ are functions which depend only on $y$. The isometries generically preserved by these backgrounds are easily seen to correspond to the symmetries characterizing the conformal monodromy defects introduced above. In particular, the AdS$_{d-1}$ foliation gives rise to the $SO(d-2,2)$ factor, while the invariance under translations along $z$ coincides with the preserved $SO(2)$. 

Consistent with these isometries, the bulk gauge fields holographically dual to the global symmetry currents coupling to the monodromy sources are taken to be
\begin{equation}
A^I = a^I \dd z
\end{equation}
with $a$ a function of $y$. There may also be other bulk fields present in the solution, such as scalars---again they are taken to depend only on the $y$-coordinate, but will play little role in what follows. 

When an explicit parametrisation is required, we take coordinates on the AdS$_{d-1}$ leaves such that
\begin{equation}
\dd s^2\left(\mathrm{AdS}_{d-1} \right) = \frac{1}{Z^2}\left(-\dd t^2 +\dd Z^2+\dd r_{||}^2 +r_{||}^2\,\dd \Omega_{d-4}^2\right),
\end{equation}
with $\Omega_{d-4}^2$ a metric on the unit radius $\mathbb{S}^{d-4}$. For the special case in which $d=3$, the AdS$_2$ leaves are parametrised by the coordinates $(t,Z)$ alone. 

As shown by \cite{Ryu:2006bv, Hubeny:2007xt}, the quantum entanglement in a holographic quantum field theory between a spherical entangling region and its complement is encoded gravitationally in the area of a particular minimal surface within the bulk. This prescription was first applied to bulk solutions holographically dual to defect conformal field theories in \cite{Jensen:2013lxa}. There it was shown that for a codimension two defect in a $d$ dimensional field theory, the spherical entanglement entropy for an entangling region of radius $R$ is given by the integral expression
\begin{equation}\label{eq:S_EEd4p}
S_{EE} = \frac{\pi}{2 G^{(d+1)}_N}\mathrm{vol}(\mathbb{S}^{d-4})\int \dd r_{||}\frac{R\, r_{||}^{d-4}}{\left(R^2-r_{||}^2 \right)^{(d-2)/2}}\int \dd y\, e^{(d-3)V} | f h |.
\end{equation}
Some care is required when using this expression to compute holographic entanglement entropies. In particular, by convention one is to take $\mathrm{vol}(\mathbb{S}^0)=2$, and moreover, when $d=3$ the correct expression reduces to
\begin{equation}\label{eq:S_EEd3}
S_{EE} = \frac{\pi}{2 G^{(4)}_N}\int \dd y\,| f h |. \qquad (d=3)
\end{equation}
One must also pay careful attention to the limits of integration. As a consequence of the cut-off procedure used to holographically regulate the short distance divergences which appear in the entanglement entropy, the integrals in (\ref{eq:S_EEd4p}) do not generically factorize. 

While a careful discussion of the details can be found in appendix B of \cite{Jensen:2013lxa}, the main idea is as follows: upon implementing a conformal transformation such that the dual dCFT inhabits $\mathbb{R}^{1,d-1}$, the bulk solution is brought to a form in which distances from the boundary are measured by a Fefferman-Graham (FG) radial coordinate which can be thought of as tracking energy scales in the dual theory in the familiar way. A short distance cut-off $\epsilon$ is then implemented on this FG radial coordinate. 

The integral over $r_{||}$ is subsequently taken over $r_{||} \in [0,r_\epsilon]$ with $r_\epsilon = \sqrt{R^2-\epsilon^2}$. The integral over $y$ also needs to be regulated in an analogous manner. In this case, however, the coordinate transformation implementing the boundary conformal transformation muddies the waters somewhat. The FG cut-off dual to the short distance regulator becomes a surface in the original coordinates which can be parametrised as $y_\epsilon(r_{||},R)$. 

As the notation suggests, this cut-off in $y$ introduces additional $r_{||}$ dependence which must be accounted for in the integral over $r_{||}$. For the bulk solutions holographically dual to codimension two defects studied in this work, the coordinate $y$ then takes values $y \in [y_c,y_\epsilon]$, where $y_c$ is the location of the holographic defect ``core''---the location in the bulk where the spatial geometry smoothly caps off in a regular way. 

\subsection{Entropic central charges for holographic defects}
As emphasised above, the defect entanglement entropy is generically a divergent quantity regulated by the short distance cut-off $\epsilon$. Thus, in order to extract physical data characterising the defect, it is first necessary to understand the structure of these divergences. This structure, in turn, depends on  both the dimensionality of the bulk CFT and the codimension of the defect it supports.

In \cite{Jensen:2013lxa} an argument is provided, based on an appeal to weak coupling intuition and the result of \cite{PhysRevLett.67.161}, that in the $\epsilon\to0$ limit the spherical entanglement entropy for the defect generically contains  both divergent and non-divergent terms characteristic of the vacuum (i.e. the bulk theory in the absence of the defect) as well as terms inherent to the presence of the defect. The former are of course agnostic to the details of the defect. Accordingly, a simple prescription for extracting physical information characterising the defect is provided---one simply computes the difference in the entanglement entropy between the dCFT and the vacuum, removing these agnostic terms:
\begin{equation}
\Delta\mathcal{S}_{EE} \equiv S^{\mathrm{dCFT}}_{EE} - S_{EE}^0.
\end{equation}

Specialising to codimension two monodromy defects, which are parametrised by monodromy parameters $g\mu^I$ and (optionally) by the presence of a conical deficit/excess at the location of the defect with parameter $n >0$, we analogously define the quantity
\begin{equation}
\Delta\mathcal{S}_{EE} = S_{EE}[g\mu^\alpha;n] - n S_{EE}[0;1],
\end{equation}
which we refer to as the ``defect entanglement entropy''. For such defects, the expected form of the defect entanglement entropy in the $\epsilon \to 0$ limit is
\begin{equation}
\Delta \mathcal{S}_{EE} = c_{d-4} \frac{R^{d-4}}{\epsilon^{d-4}}+
\begin{cases}
\mathcal{C}_\mathcal{D}^{(d)} + \ldots & \qquad\qquad  d \quad \mathrm{odd}\\
& \mathrm{for}\\
\mathcal{C}_\mathcal{D}^{(d)}\log\left(\frac{2R}{\epsilon} \right) + \ldots & \qquad\qquad d >2 \quad \mathrm{even}
\end{cases}
\end{equation}
where the constants $c_{d-4}$ are non-vanishing only for $d\ge5$. Unlike the $c_{d-4}$, the coefficients $\mathcal{C}_\mathcal{D}^{(d)}$ are cut-off independent---rescaling the short distance cut-off $\epsilon$ does not modify their value. In this sense they are sometimes referred to as ``universal'' and can be used to quantify physical properties of the defect.

In the {\it absence} of defects, the vacuum analogues of the $\mathcal{C}_\mathcal{D}^{(d)}$---which we might refer to as $\mathcal{C}^{(d)}$---appearing in the entanglement entropy play a distinguished role  \cite{Klebanov:2011gs,Jafferis:2011zi,Myers:2010tj,Casini:2011kv,Casini:2012ei}. For example, when $d = 3$ the coefficient $\mathcal{C}^{(3)}$ is be proportional to the partition function of the Euclidean CFT$_3$ on $\mathbb{S}^3$. In $d=4$ on the other hand, $\mathcal{C}^{(4)}$ is proportional to the $a$ central charge which appears in front of the Euler density term in the trace anomaly. Importantly, it is generally expected (and in some cases proven to be true) that the $\mathcal{C}^{(d)}$ obey monotonicity properties along an RG flow, decreasing from UV to IR. For this reason, the $\mathcal{C}^{(d)}$ are sometimes referred to as ``entropic $c$-functions'' \cite{Liu:2012eea}. It is an open and interesting question whether or not the $\mathcal{C}_\mathcal{D}^{(d)}$ obey similar monotonicity properties. This question largely motivates the present work.

For the codimension two monodromy defects discussed in this work, it turns out that one can compute the $\mathcal{C}_\mathcal{D}^{(d)}$ fairly simply. Let us first introduce the notation $[\ldots]_\mathcal{D}$, where for any defect quantity $\mathcal{F}$ depending on the defect data, $\mathcal{F}[g\mu^I;n]$, we have
\begin{equation}
\left[\, \mathcal{F}\,\right]_\mathcal{D} \equiv \mathcal{F}[g\mu^I;n]-n\,\mathcal{F}[0,1].
\end{equation}
We then find that for the class of defects studied here,
\begin{equation}\label{eq:Ceq}
\mathcal{C}_\mathcal{D}^{(d)} = -2^{\lceil\frac{d}{3}\rceil}\frac{(-\pi)^{\lceil\frac{d}{4} \rceil}}{4G_N^{(d+1)}}\left[ \int \dd y\, e^{(d-3)V} | f h |\right]_\mathcal{D} \qquad \mathrm{for}\quad 3\le d \le 6
\end{equation}
with $\lceil . \rceil$ the ceiling function. As before, the $y$-integral is over $y \in [y_c,y_\epsilon]$, as explained in further detail in appendix \ref{app:cut}. 

Our primary aim is two-fold: we wish to evaluate (\ref{eq:Ceq}) explicitly for families of superconformal monodromy defects in well known holographic field theories, and subsequently investigate the behavior of $\mathcal{C}_\mathcal{D}^{(d)}$ under renormalisation group flow. 

\section{Defect entanglement entropy in ABJM theory}\label{sec:3d}
A simple class of holographic superconformal monodromy defects in $d=3$ are provided by the solutions to the maximal $SO(8)$ gauged supergravity in $D=4$ recorded in  (\ref{eq:4dU14solution}). These solutions can be uplifted on the seven sphere to $D=11$ supergravity \cite{Cvetic:1999xp}, and are dual to conformal monodromy defects in the ABJM theory which generically preserve two Poincar\'e supercharges. In addition to the conical singularity parameter $n$, the defects are characterized by three independent flavor monodromy sources.
\subsection{Results}
Evaluating (\ref{eq:Ceq}) on the holographic monodromy defect solutions within the STU truncation of $D=4$ gauged supergravity (\ref{eq:4dU14solution}) yields
\begin{equation}\label{eq:4dgen}
\mathcal{C}_\mathcal{D}^{(3)} = \frac{\pi L^2}{4 G_N^{(4)}}\left[ n\int\mathrm{d}y\right]_\mathcal{D} = \frac{\pi L^2}{4 G_N^{(4)}}n\left(y_\epsilon-y_\epsilon^0 + y_c^0-y_c \right)
\end{equation}
where the superscript ``$\,\,^0\,$'' denotes a quantity evaluated in the background dual to the (defectless) vacuum. We thus see that the crux of the computation becomes the evaluation of the quantities $y_\epsilon$ and $y_c$ in terms of the defect parameters $n$ and $g\mu^I$.

While the general result written in terms of arbitrary values for the flavor monodromies is rather unwieldy, there are simple subtruncations of the STU theory that provide illustrative examples. We first exhibit one such example, before turning to the general case.

Notably, the result for the minimal gauged supergravity defect (\ref{eq:functions4Dmin}) can be written
\begin{equation}
\mathcal{C}_\mathcal{D}^{(3)}= \frac{\pi L^2}{4G_N^{(4)}}\left(n-1\right) = \frac{1}{2}F^{ABJM}_{S^3}\left(n-1\right)
\end{equation}
where 
\begin{equation}
F^{ABJM}_{S^3} = \frac{\pi L^2}{2 G_N^{(4)}} = \frac{\sqrt{2}\pi}{3} N^{3/2}
\end{equation}
is the free energy of ABJM on $S^3$. As the monodromy defects dual to the minimal gauged supergravity solutions are necessarily accompanied by a conical singularity at the location of the defect, it is reassuring that $\mathcal{C}_\mathcal{D}^{(3)}$ vanishes when $n=1$ in this case.

Interestingly, using the results of \cite{Arav:2024wyg}, we observe that this defect entanglement entropy can be written in terms of the defect field theory data $\mathcal{I}_D$ and $h_D$, where
\begin{equation}
\mathcal{I}_D\equiv -\Big[I_n[g\mu^I]\Big]_\mathcal{D}
\end{equation} and $I_n[g\mu^I]$ is the Euclidean on-shell action of the dCFT. In particular, in our conventions we have
\begin{equation}
h_D = \frac{\pi}{2n}\sum_I \langle J^I \rangle \qquad \mathrm{with} \qquad \langle J^I \rangle = -\frac{N^{3/2}}{6\sqrt{2}\pi}\left(\frac{2g\mu^I}{1+\frac{2g\mu^I}{n} }\right)\mathcal{F}^{ABJM}
\end{equation}
and
\begin{equation}
\mathcal{F}^{ABJM} \equiv \left[ \left(1+\frac{2\,g\mu_0}{n}  \right)\left(1+\frac{2\,g\mu_1}{n}  \right)\left(1+\frac{2\,g\mu_2}{n}  \right)\left(1+\frac{2\,g\mu_3}{n}  \right)\right]^{1/2},
\end{equation}
as well as
\begin{equation}
\mathcal{I}_\mathcal{D} = n\left(1-\mathcal{F}^{ABJM} \right)F_{S^3}^{ABJM}.
\end{equation}

Importing these results, we eventually find
\begin{equation}\label{eq:dS4d}
\mathcal{C}_\mathcal{D}^{(3)}=\mathcal{I}_D-2\pi n\,h_D.
\end{equation}
In fact, this expression holds not only for the minimal gauged supergravity defect, but for the {\it entire class} of STU superconformal monodromy defects considered here!

Such a relationship is not entirely unexpected. Indeed, the $n\to1$ limit of the supersymmetric charged Renyi entropy computed in \cite{Arav:2024wyg} can be used to define a defect entanglement entropy for monodromy defects {\it without} conical singularities. The supersymmetric charged Renyi entropy is defined as
\begin{equation}
\mathcal{S}_n = \frac{I_n[g\mu^I]-n I_{n=1}[g\mu^I]}{n-1},
\end{equation}
with all supersymmetry constraints on the monodromy sources imposed. As explained in \cite{Arav:2024wyg}, some care is required in taking the $n\to1$ limit---here we will take the limit by keeping the independent flavor monodromy sources defined in (\ref{eq:mu4d}) fixed.

Introducing the quantity
\begin{equation}
\Delta \mathcal{S}_1 \equiv \mathcal{S}_1 -\mathcal{S}_1\big |_{g\mu_F=0},
\end{equation}
which is the difference between the $n\to 1$ limit of the supersymmetric charged Renyi entropy for arbitrary flavor monodromies and for the ABJM vacuum, explicit computation yields
\begin{equation}\label{eq:dSRenyi4D}
\Delta\mathcal{S}_1=\mathcal{I}_D-2\pi\,h_D
\end{equation}
which agrees with (\ref{eq:dS4d}) evaluated at $n=1$.

\subsection{RG flows and $\mathcal{C}_\mathcal{D}^{(3)}$}\label{sec:RG4d}
In conformal field theories without defects, the scheme independent part of the entanglement entropy has been shown to exhibit monotonicity along a renormalisation group flow. It is interesting to wonder whether this continues to be the case for defect conformal field theories.

Towards this end, we can evaluate the defect entanglement entropy (\ref{eq:dS4d}) at the endpoints\footnote{Strictly speaking, we have not shown here that $\mathcal{C}_\mathcal{D}^{(3)}$ for the monodromy defect in mABJM theory takes the form (\ref{eq:dS4d}), though for $n=1$ evidence for this is provided by a similar analysis to that which led to (\ref{eq:dSRenyi4D}).}  of a putative RG flow between a monodromy defect in the ABJM and mABJM theories. We first motivate the existence of such a flow.  In the absence of defects, a supersymmetric RG flow can be induced by deforming the ABJM theory (in flat space) via a spatially homogenous mass term for a particular combination of bosonic and fermionic bilinears \cite{Corrado:2001nv,Benna:2008zy}. 

In the presence of defects, it is thus natural to posit the existence of an RG flow from the ABJM defect with $\mu_B=0$, and the mABJM defect with the {\it same} values of $(n,\mu_{F_1},\mu_{F_2})$. Such a flow would be driven by the same (homogenous) supersymmetric mass deformation---in other words it is an RG flow driven by an operator which is not localised on the defect. Whether or not such an RG flow exists, even in the holographic limit, is a difficult question to fully address. As emphasised in \cite{Arav:2024wyg}, there exist ABJM defects with $\mu_B=0$ and values of $(n,\mu_{F_1},\mu_{F_2})$ for which {\it no} mABJM defect can be realised. In what follows, we will restrict our attention to putative flows between ABJM and mABJM defects with $(n,\mu_{F_1},\mu_{F_2})$ values that satisfy the naive existence bounds, keeping in mind that the allowed parameter space of flows is not yet known.

\begin{figure}[h]
\centering
\includegraphics[scale=0.46]{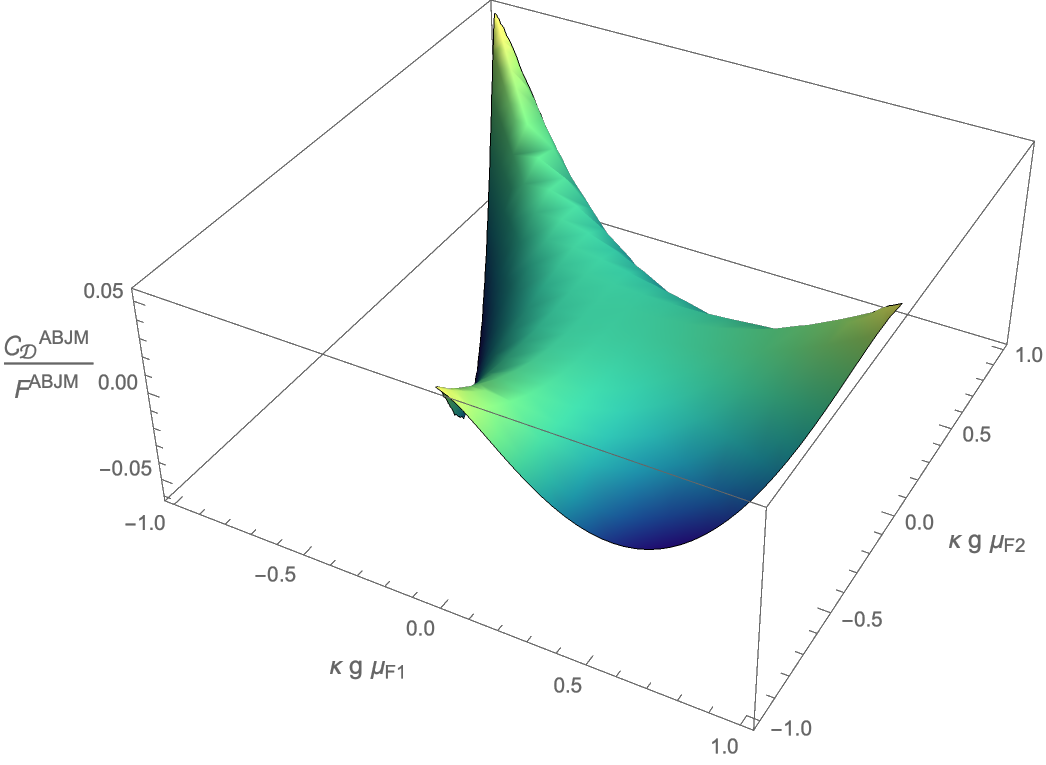}
\includegraphics[scale=0.4]{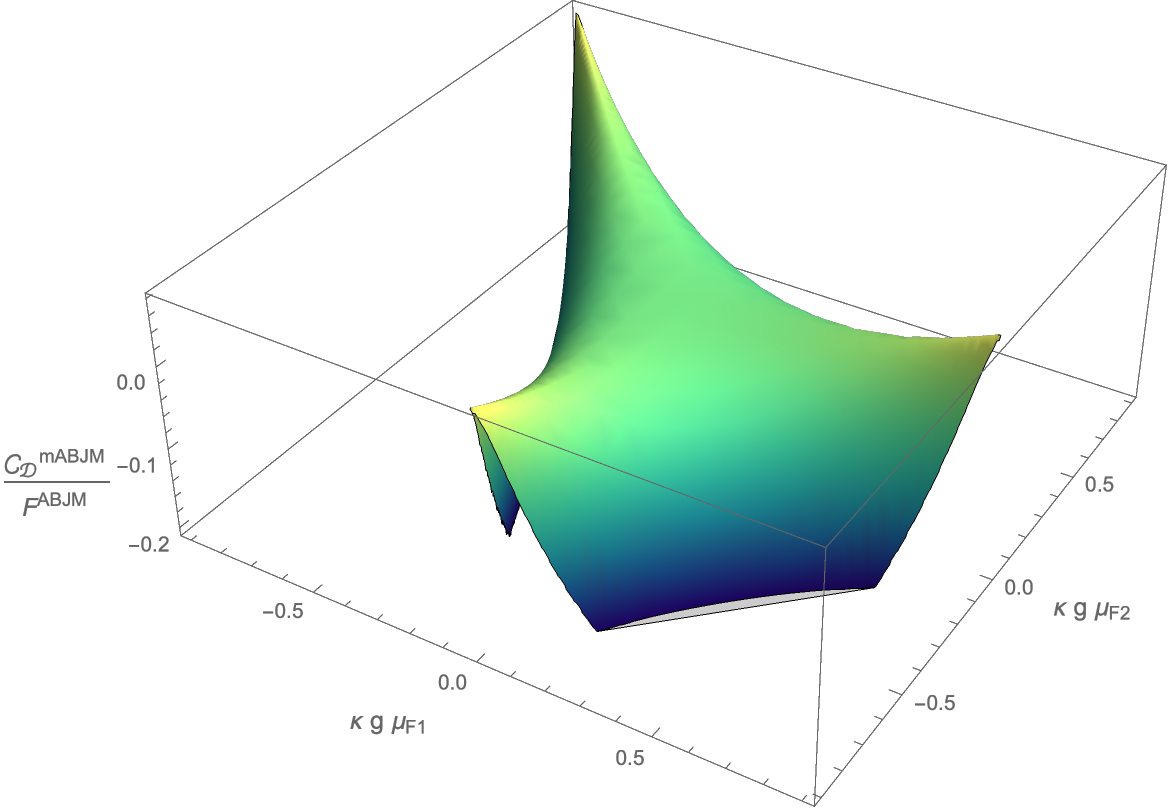}
\caption{The scheme independent part of the defect entanglement entropy, $\mathcal{C}_\mathcal{D}^{(3)}$, for ABJM with $\mu_B=0$ (left) and mABJM (right) defects with $n=1$. The parameter $\kappa = \pm1$ appears here to aid in comparison with \cite{Arav:2024wyg}---in this work we take $\kappa = 1$ without loss of generality.}\label{fig:CD4Dnis1}
\end{figure}

One may gain some intuition for the behaviour of the defect entanglement entropy along such an RG flow by considering a flow between defects absent of conical singularities (i.e. with $n=1$). The quantity $\mathcal{C}_\mathcal{D}^{(3)}$ with $n=1$ is plotted for both the ABJM and mABJM defect theory in figure \ref{fig:CD4Dnis1}. It is immediately clear that within the ``allowed range" of monodromy parameters $(\mu_{F_1},\mu_{F_2})$ for the putative flow, in both cases $\mathcal{C}_\mathcal{D}^{(3)}$ can vanish, and indeed change sign. Such behaviour has been noted previously, for example in \cite{Capuozzo:2023fll}.

\begin{figure}[h]
\centering
\includegraphics[scale=0.4]{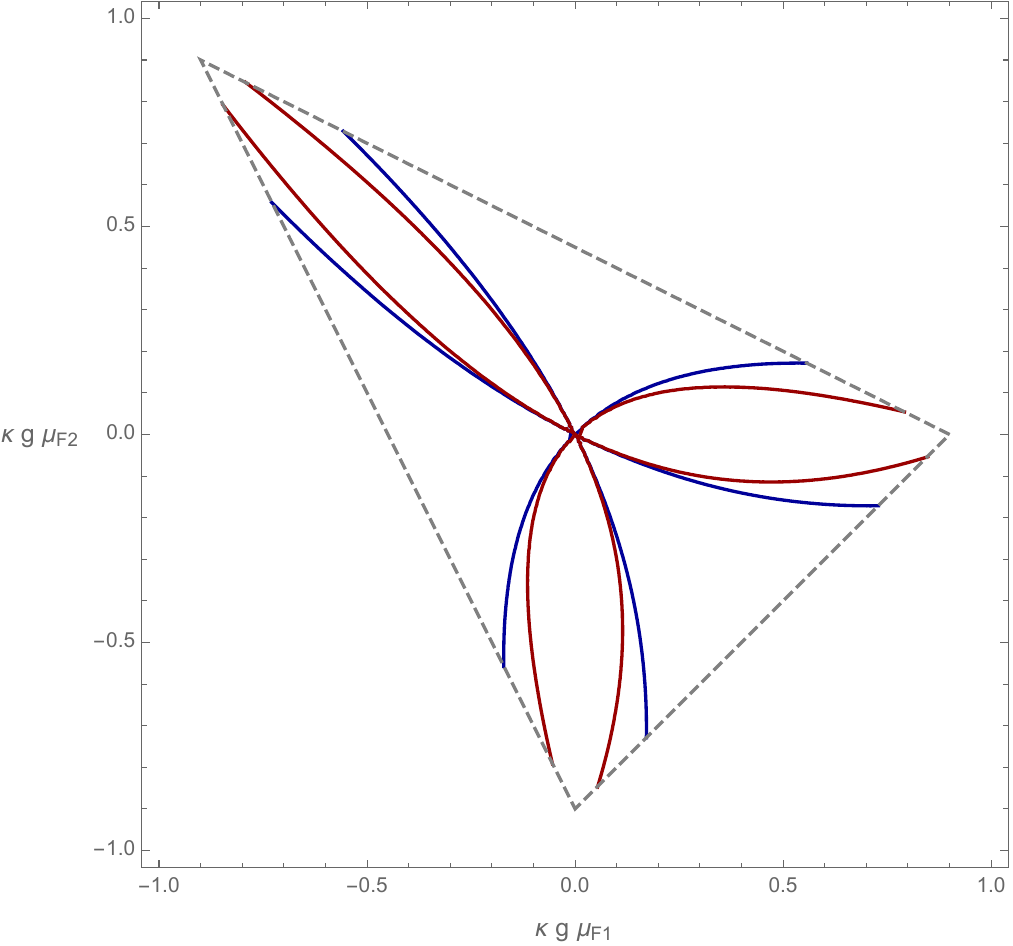}
\caption{The locus of monodromy parameters at which $\mathcal{C}_\mathcal{D}^{(3)}$, for ABJM with $\mu_B=0$ (blue) and mABJM (red) defects with $n=1$ vanishes. The ``allowed region'' of monodromy parameters under which a flow naively might exist is given by the interior of the dashed triangle.}\label{fig:C4dzeroLoc}
\end{figure}

The locus of monodromy parameters which lead to a vanishing $\mathcal{C}_\mathcal{D}^{(3)}$ in the absence of conical deficits is shown in figure \ref{fig:C4dzeroLoc}. Importantly, one observes that this locus does not everywhere coincide for the ABJM and mABJM defects. It is thus immediately clear that in the absence of conical deficit, the defect entanglement entropy, $\mathcal{C}_\mathcal{D}^{(3)}$, can not be monotonic along such bulk RG flows.

\begin{figure}[h]
\centering
\includegraphics[scale=0.4]{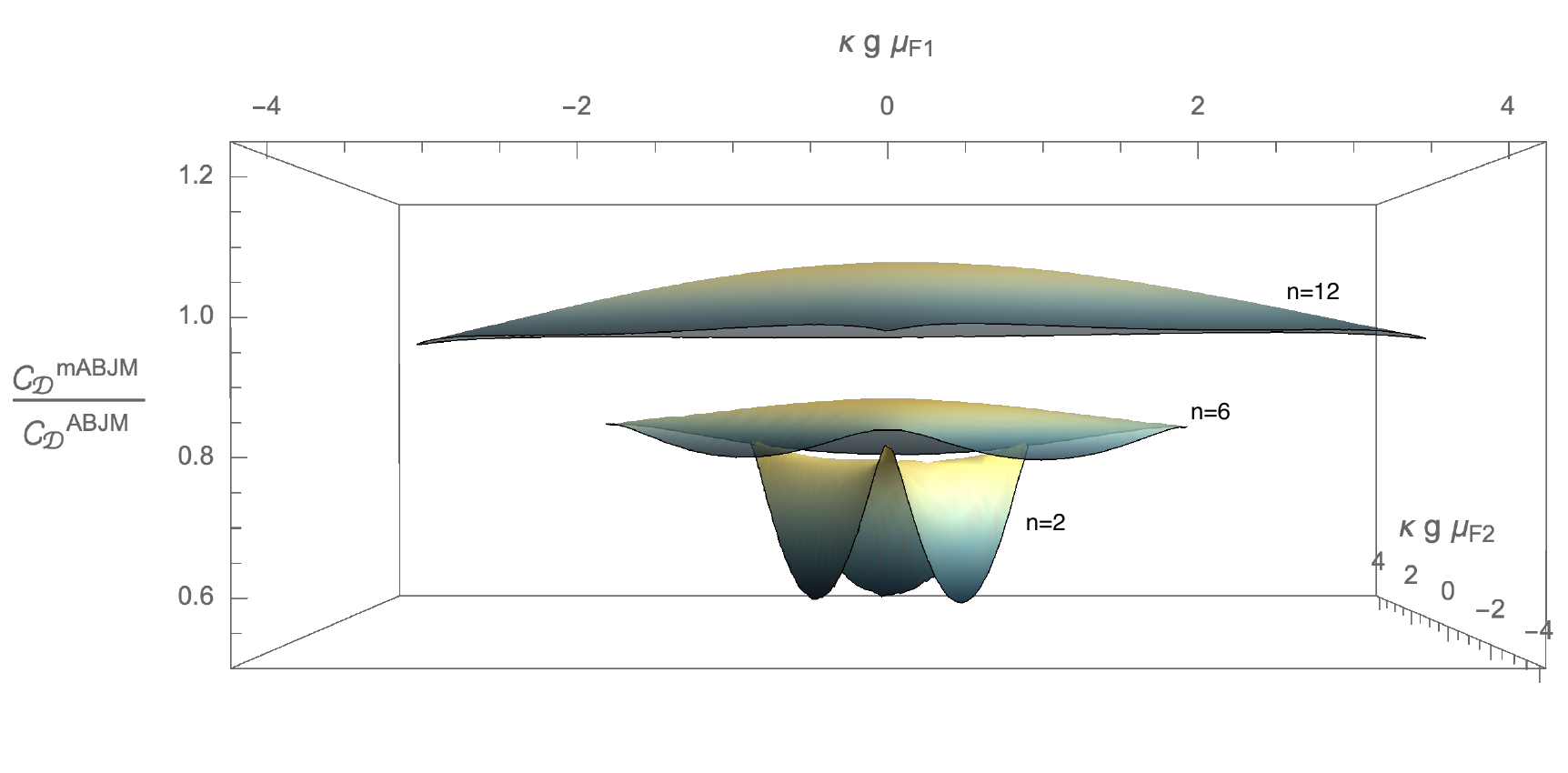}
\caption{The ratio of $\mathcal{C}_\mathcal{D}^{(3)}$ between mABJM and ABJM (with $\mu_B=0$)  defects for various values of $n>1$. The monotonicity of this quantity depends on the value of $n$ characterising the conical deficit.}\label{fig:CD4Drat}
\end{figure}

Generalizing to flows between defects with $n >1$, one finds that monotonicity of $\mathcal{C}_\mathcal{D}^{(3)}$ is not generic, but indeed depends on the value of $n$. This is illustrated for several representative examples in figure \ref{fig:CD4Drat}. There, the ratio of $\mathcal{C}_\mathcal{D}^{(3)}$ between the ABJM and mABJM defects is given. Empirically, one finds that for sufficiently large---but not too large---values of $n$ (e.g. $n =12$), the quantity $\mathcal{C}_\mathcal{D}^{(3)}$ may either increase or decrease from UV to IR. 

Moreover, we note that the value this ratio obtains as one removes the defect depends on the way in which this limit is taken. Specifically, we find that 
\begin{equation}
\frac{\mathcal{C}_\mathcal{D}^{(3)}[\mathrm{ABJM}]\big|_{g\mu_B=0}}{\mathcal{C}_\mathcal{D}^{(3)}[\mathrm{mABJM}]} = 
\begin{cases}
\frac{2}{3\sqrt{3}} & \qquad \mathrm{for} \qquad n =1, \quad \mu_{F_i}\to 0\\
&\\
\frac{3\sqrt{3}}{4} =  \frac{F^{ABJM}_{S^3}}{F^{mABJM}_{S^3}}& \qquad \mathrm{for} \qquad \mu_{F_i}=0, \quad n\to 1
\end{cases}.
\end{equation}
Thus, although the defect entanglement $\mathcal{C}_\mathcal{D}^{(3)}$ vanishes in the absence of defects independent of the order of limits taken, the ratio of this quantity in the presumed UV and IR remains finite. Indeed, absent flavor monodromies, as the conical deficit is removed the ratio attains the value $\mathcal{C}^{(3)}[\mathrm{ABJM}]/\mathcal{C}^{(3)}[\mathrm{mABJM}]$, which is to say the value of the ratio of the scheme independent parts of sphere entanglement entropy in the (defectless) UV and IR vaccua.

\section{Defect entanglement entropy in $\mathcal{N}=4$ SYM theory}\label{sec:4d}
A class of holographic superconformal monodromy defects in four dimensions is provided by the solutions to $SO(6)$ gauged supergravity in $D=5$ recorded in (\ref{eq:5DU13solution}). These solutions can be uplifted on the five sphere to type IIB supergravity \cite{Cvetic:1999xp}, and are holographically dual to two dimensional $\mathcal{N}=(0,2)$ superconformal defects in $\mathcal{N}=4$ Super Yang-Mills theory. In addition to the conical singularity parameter $n$, the defects are parametrized by two independent flavor monodromy sources. 
\subsection{Results}
Returning again to (\ref{eq:Ceq}), we find that for the holographic monodromy defect solutions within the STU truncation of $D=5$ gauged supergravity (\ref{eq:5DU13solution}) one obtains
\begin{equation}\label{eq:5dgen}
\mathcal{C}_\mathcal{D}^{(4)} = \frac{\pi L^3}{2G_N^{(5)}}\left[ n\int\mathrm{d}y\right]_\mathcal{D} = \frac{\pi L^3}{2 G_N^{(5)}}n\left(y_\epsilon-y_\epsilon^0 + y_c^0-y_c \right)
\end{equation}
where, as above, the superscript ``$\,\,^0\,$'' denotes a quantity evaluated in the background dual to the (defectless) vacuum. This expression is very similar to that of (\ref{eq:4dgen}), and like there, we see that for these solutions its evaluation simply requires understanding the dependence of the limits of integration on the defect parameters.

It is illustrative to consider first the defect entanglement entropy for the comparatively simple holographic solution to the minimal gauged supergravity theory, given by (\ref{eq:5DU13solution}) with  (\ref{eq:functions5Dmin}). One obtains 
\begin{equation}
\mathcal{C}_\mathcal{D}^{(4)}  = \frac{\pi L^3}{18 G_N^{(5)}}\frac{(n-1)(1+5n)}{n} = \frac{4(n-1)(1+5n)}{9n}a^{\mathcal{N}=4}
\end{equation}
where
\begin{equation}
a^{\mathcal{N}=4} = \frac{\pi L^3}{8 G_N^{(5)}} = \frac{1}{4}N^2
\end{equation}
is the central charge appearing in front of the Euler density term in the trace anomaly of $\mathcal{N}=4$ SYM. Again, this defect contribution to the entanglement entropy vanishes when $n=1$, as in this case there is no monodromy defect without a conical singularity in the boundary theory.

As in section \ref{sec:3d}, it is interesting to try to write $\mathcal{C}_\mathcal{D}^{(4)}$ in terms of defect quantities. In this case, the analogous quantities are found in $\cite{Arav:2024exg}$ to be
\begin{equation}
h_D = -\frac{2\pi}{3 n}\left(\langle J^1 \rangle+\langle J^2 \rangle+\langle J^3 \rangle \right)
\end{equation}
with
\begin{align}
\langle J^1 \rangle & = \frac{N^2}{4\pi^2}\left( g\mu_1\right)\left(1+\frac{g\mu_2}{n} \right)\left(1+\frac{g\mu_3}{n} \right)\nonumber\\
\langle J^2 \rangle & = \frac{N^2}{4\pi^2}\left( g\mu_2\right)\left(1+\frac{g\mu_1}{n} \right)\left(1+\frac{g\mu_3}{n} \right)\nonumber\\
\langle J^3 \rangle & = \frac{N^2}{4\pi^2}\left( g\mu_3\right)\left(1+\frac{g\mu_1}{n} \right)\left(1+\frac{g\mu_2}{n} \right)
\end{align}
as well as
\begin{equation}
b = 12 a^{\mathcal{N}=4}\, n \left(1-\mathcal{F}^{\mathcal{N}=4} \right),
\end{equation}
where
\begin{equation}
\mathcal{F}^{\mathcal{N}=4} \equiv  \left(1+\frac{g\mu_1}{n} \right)\left(1+\frac{g\mu_2}{n} \right)\left(1+\frac{g\mu_3}{n} \right).
\end{equation}

For the general case with arbitrary allowed values of the monodromy parameters, we find the notable result
\begin{equation}\label{eq:dS5d}
\mathcal{C}_\mathcal{D}^{(4)} = \frac{1}{3}\big(b - 6\pi n\, h_D \big).
\end{equation}
When $n=1$, this expression is in line with the general expectations for codimension two superconformal defects in four dimensions \cite{Jensen:2018rxu}, and is in fact compatible with the discussion in \cite{Arav:2024exg} which generalises these expectations to the case with conical deficits present.

\subsection{RG flows and $\mathcal{C}_\mathcal{D}^{(4)}$}\label{sec:RG5d}
In the spirit of section \ref{sec:RG4d}, it is natural to wonder whether or not the quantity obtained in (\ref{eq:dS5d}) is strictly decreasing along RG flows induced by operators not confined to the location of the defect. Towards this end, we consider the RG flow in four dimensions between the $\mathcal{N}=4$ SYM theory and the $\mathcal{N}=2$ Leigh-Strassler (LS) superconformal field theory \cite{Leigh:1995ep,Freedman:1999gp}. This Poincar\'e invariant flow is induced by a supersymmetric mass term for boson and fermion bilinears, and in some ways can serve as the four dimensional analogue of the flow studied in section \ref{sec:RG4d}.

Superconformal monodromy defects in the LS theory were studied holographically in \cite{Arav:2024exg}. As before, we posit the existence of an RG flow between monodromy defects in an $\mathcal{N}=4$ SYM bulk  and those in LS, induced by the same homogenous mass deformation that drives the Poincar\'e invariant flow. Should such a flow exist, we expect that the values of the monodromy sources agree in the UV and IR. In particular, one should arrange that $\mu_B=0$ also in the  $\mathcal{N}=4$ SYM defect theory.

As in section \ref{sec:RG4d}, we proceed with the caveat that we have not explicitly shown here that the defect entanglement entropy in the LS theory takes the form (\ref{eq:dS5d}). Here we leave a detailed investigation of this to the future, and note that evidence in support of (\ref{eq:dS5d}) for the LS defects is provided by the general arguments of  \cite{Jensen:2018rxu}.

Unlike the putative flows in 3 dimensional defect CFTs studied in section \ref{sec:RG4d}, here the scheme independent part of the defect entanglement entropy $\mathcal{C}_\mathcal{D}^{(4)}$ in the UV and IR theories is parametrised only by $(n, \mu_F)$. In fact, it can easily be written in closed form:
\begin{equation}
\mathcal{C}_\mathcal{D}^{(4)}[\mathcal{N}=4]\big|_{g\mu_B=0} = \frac{a^{\mathcal{N}=4}}{12 n}\left(4 (g\mu_F)^2 +27 n^2 -22 n -5\right)
\end{equation}
and
\begin{equation}
\mathcal{C}_\mathcal{D}^{(4)}[\mathrm{LS}]= \frac{a^{\mathcal{N}=4}}{8 n}\left(4 (g\mu_F)^2 +15 n^2 -12 n -3\right).
\end{equation}
An immediate observation is that the ratio
\begin{equation}
\frac{\mathcal{C}_\mathcal{D}^{(4)}[\mathcal{N}=4] \big|_{g\mu_B=0}}{\mathcal{C}_\mathcal{D}^{(4)}[\mathrm{LS}]} = 
\begin{cases}
\frac{2}{3} & \qquad \mathrm{for} \qquad n =1, \quad \mu_F\to 0\\
&\\
\frac{32}{27} =  \frac{a^{\mathcal{N}=4}}{a^{\mathrm{LS}}}& \qquad \mathrm{for} \qquad \mu_F=0, \quad n\to 1
\end{cases}
\end{equation}
and thus, as before, the value of this ratio along such a flow can depend on the way a point in the parameter space is approached. In particular, for defects without conical singularity the quantity $\mathcal{C}_\mathcal{D}^{(4)}$ is seen to {\it increase} along the flow by a factor of 3/2 across the allowed region, $|g\mu_F|<3/2$. Conversely, in the absence of flavor monodromies, as one removes the conical deficit this ratio reduces to precisely the value obtained in the homogenous RG flow $a^{\mathcal{N}=4}/a^{LS}$.

\begin{figure}[h]
\centering
\includegraphics[scale=0.6]{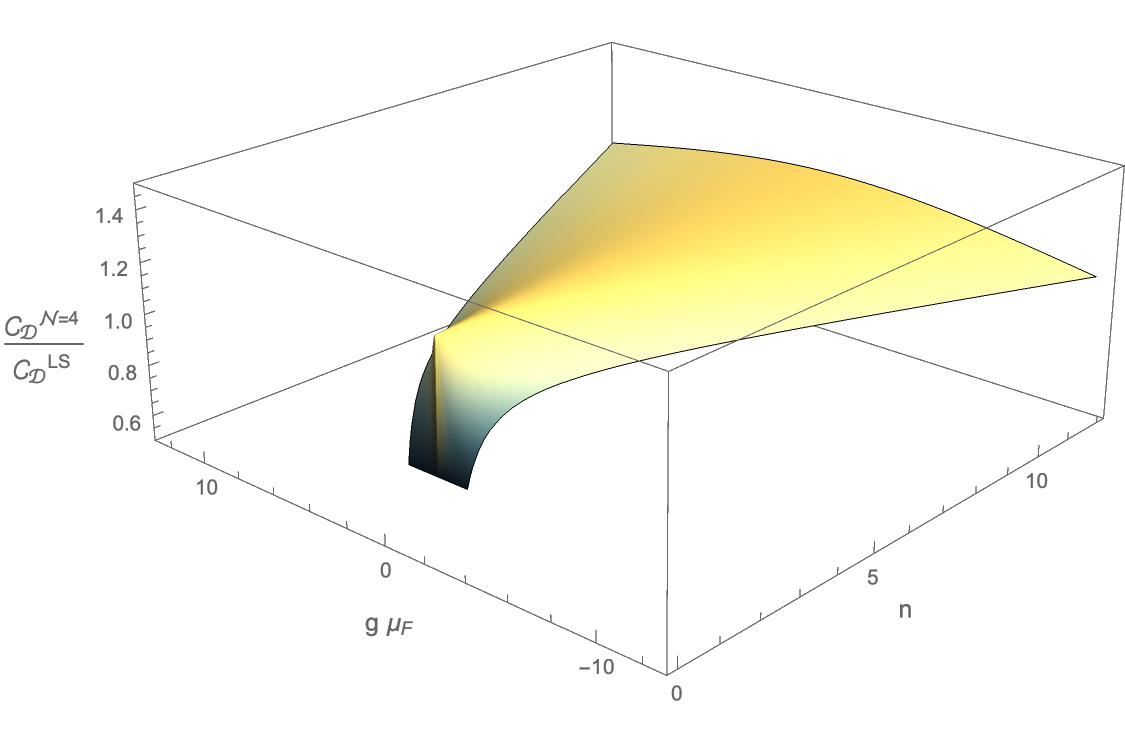}
\caption{The ratio of $\mathcal{C}_\mathcal{D}^{(4)}$ between $\mathcal{N}=4$ (with $g\mu_B = 0$) and LS defects  for various values of $n\ge1$. }\label{fig:CD5Drat}
\end{figure}

The behaviour of this ratio throughout the naively allowed parameter space is shown in figure \ref{fig:CD5Drat}. Interestingly, only for sufficiently large values of $n$ is this ratio greater than one across the entirety of the allowed region. 

\section{Defect entanglement entropy in the $d=6$ $\mathcal{N}=(2,0)$ theory}\label{sec:6d}
The defect entanglement entropy for codimension two monodromy defects in the $d=6$ $\mathcal{N}=(2,0)$ theory was discussed in detail in \cite{Capuozzo:2023fll}. Here\footnote{In the section, the conventions for $h_D$ differ slightly from those of \cite{Capuozzo:2023fll}, to remain consistent with the definitions used in sections \ref{sec:3d} and \ref{sec:4d} of this work.} we generalise those results to include non-trivial deficit angles in the boundary theory, and broaden slightly the holographic perspective on these systems. The conformal monodromy defects discussed here preserve $\mathcal{N}=1$ supersymmetry in four dimensions, and are generically characterized by a conical singularity parameter $n$ and a single independent flavor monodromy source.
\subsection{Results}
Once more appealing to (\ref{eq:Ceq}), we find that for the holographic monodromy defect solutions within the STU truncation of $D=7$ gauged supergravity (\ref{eq:D7U12solution}) one obtains
\begin{equation}
\mathcal{C}_\mathcal{D}^{(6)} = -\frac{\pi ^2 L^5}{32G_N^{(7)}}\left[ n\int y\,\mathrm{d}y\right]_\mathcal{D} = -\frac{\pi^2}{64}\frac{L^5}{G_N^{(7)}}n\Big(y_\epsilon^2-(y_\epsilon^0)^2 - y_c^2+(y_c^0)^2 \Big)
\end{equation}
As previously, the superscript ``$\,\,^0\,$'' denotes a quantity evaluated in the background dual to the (defectless) vacuum. Unlike (\ref{eq:4dgen}) and (\ref{eq:5dgen}), in these coordinates the expression for the defect entanglement entropy does not evaluate to a total integral. Importantly however, from the results in appendix \ref{app:cut} we find that upon proper application of the vacuum subtraction prescription it remains both finite and independent of $r_{||}$.

The defect entanglement entropy for these holographic monodromy defects is found to be
\begin{multline}
\mathcal{C}_\mathcal{D}^{(6)}=-a^{(2,0)}\frac{\pi^3}{3840 n^3}\Big((n-1)\left(1+n\left(29+n\left(227+767 n \right) \right) \right)\\
+ 2(1+3n)(1+11n)(g\mu_F)^2-(g\mu_F)^4\Big)
\end{multline}
where $a^{(2,0)}$ is the $a$ central charge of the 6d theory, given in the conventions of  \cite{Henningson:1998gx} as
\begin{equation}\label{eq:a6d}
a^{(2,0)} = \frac{3}{4\pi}\frac{L^5}{G_N^{(7)}} = \frac{4}{\pi^3}N^3.
\end{equation}
As written, it is easy to see that $\mathcal{C}_\mathcal{D}^{(6)}$ vanishes in the absence of defects, which is to say when $n=1$ and $\mu_F=0$. 

From the analysis of \cite{Chalabi:2021jud}, it is known that the defect entanglement entropy in this theory takes the form
\begin{equation}\label{eq:cD7D}
\mathcal{C}_\mathcal{D}^{(6)}= -\left( 4 \mathcal{A}_\mathcal{D}-n\,\pi^2\, h_D\right)
\end{equation}
where we have generalised their results to theories with conical deficit, i.e. $n \ge 1$. Here $\mathcal{A}_\mathcal{D}$ is the defect central charge governing the Euler density term in the defect trace anomaly.

To compute the defect conformal weight $h_D$, one may use the fact that the expectation value of the renormalized stress tensor (in the conventions of \cite{Bianchi:2021snj}) for codimension two conformal defects in $d=6$ can be written
\begin{equation}
\langle \mathcal{T}_{ab} \rangle\dd x^a \dd x^b = -\frac{h_D}{2\pi}\left[\dd s^2(AdS_5)-5 n^2 \dd z^2 \right].
\end{equation}
In order to compute this expectation value holographically, we exploit the full power of holographic renormalisation \cite{deHaro:2000vlm} in a manner analogous to that used in \cite{Arav:2024wyg,Arav:2024exg}. This analysis is somewhat technical, and relegated to the appendix in \ref{sec:appOS}. There we also provide evidence that $h_D$ computed in this way is consistent with expectations imposed by supersymmetry.

In this way, we compute
\begin{equation}\label{eq:hD6d}
h_D = -a^{(2,0)}\frac{\pi}{960 n^4}\big(1+3n-g\mu_F \big)\big(1+3n+g\mu_F \big)\big(1-n(3n-2)-(g\mu_F)^2 \big)
\end{equation}

and from (\ref{eq:cD7D}) we then immediately obtain
\begin{equation}
\mathcal{A}_\mathcal{D} = -a^{(2,0)}\frac{\pi^3}{3072 n^3}\left(1+n(25n+6)-(g\mu_F)^2 \right)\left(1-n(7n-6)-(g\mu_F)^2\right).
\end{equation}

The supersymmetry algebra places strong constraints on the expectation values of various operators in the defect conformal field theory. In particular, generalising trivially the results of \cite{Bianchi:2019sxz} it is known that for superconformal monodromy defects in $d=4$ theories on AdS$_3 \times \mathbb{S}^1_n$ which preserve $\mathcal{N}=(0,2)$ supersymmetry, the defect conformal weight is proportional to the $R$-symmetry one point function. This result was realised holographically for the superconformal defects studied in \cite{Arav:2024exg}, and an analogous relation was shown to hold also for the three dimensional superconformal defects of \cite{Arav:2024wyg}.

The analysis in appendix \ref{app:cut} reveals that the superymmetric monodromy defects studied in this section are characterised by one point functions which obey a similar relation. In particular, in our conventions we have
\begin{equation}\label{eq:JRhD}
\langle J^R\,_z \rangle \equiv \langle J^R \rangle =n\frac{5 }{8\pi}h_D.
\end{equation}
with $h_D$ given by (\ref{eq:hD6d}).

Moreover, we find that the one point function of the flavor current which couples to the flavor monodromy source $\mu_F$ is given by
\begin{equation}
\langle J^F\,_z \rangle \equiv \langle J^F \rangle = a^{(2,0)}\frac{1}{384n^2}g\mu_F\left((1+3n)^2-(g\mu_F)^2 \right).
\end{equation}
Interestingly, using these definitions we observe that the differential relation
\begin{equation}
\dd \mathcal{A}_\mathcal{D} = \frac{\pi^3}{2}\left[\frac{1}{n}\langle J^F \rangle\, \dd[g\mu_F] -\frac{3}{n}\langle J^R\rangle\,\dd n\right] + \frac{\pi^3}{12} a^{(2,0)} \, \dd n
\end{equation}
holds, broadly in-line with general expectation for conformal monodromy defects \cite{Bianchi:2021snj}, generalised to the situation in which conical deficits are present. 

\section{Discussion}\label{sec:dis}
In this work, we have used gauge/gravity duality to compute the defect entanglement entropy for simple examples of superconformal monodromy defects in $d=3,4$ and 6. By combining the results of this computation with those of other observables characterizing these defect theories, we have demonstrated explicitly in each case that the defect entanglement entropy can be written in terms of the defect conformal weight $h_D$ and a dimension dependent quantity characteristic of the defect. In three dimensions this quantity is $\mathcal{I}_D$, the ``defect free energy'' computed for these defects in \cite{Arav:2024wyg}. In dimensions four and six, this quantity is the defect conformal anomaly proportional to the Euler density on the defect worldvolume.

Central to our analysis is the holographic formula (\ref{eq:Ceq}) which provides a concise means by which one may extract the ``universal'' part of the defect's contribution to the spherical entanglement entropy for these monodromy defects. It is interesting to compare this formula to other holographic ``central charge'' formulae. Generalizing slightly the discussion in appendix E of \cite{Gauntlett:2005ww}, we consider a (super)gravity theory in $D$ dimensions, with action
\begin{equation}\label{eq:SD}
S=\frac{1}{2\kappa_D^2}\int\mathrm{d}^Dx\sqrt{-G}R[G]+\ldots
\end{equation} 
It is straightforward to show that for any warped product manifold of the form
\begin{equation}
\mathrm{d}s_D^2 = e^{2\phi_1}\mathrm{d}s_q^2+e^{2\phi_2}\mathrm{d}s_p^2
\end{equation}
with Lorentzian $\mathrm{d}s^2_q$ and Euclidean $\mathrm{d}s^2_p$ and where $p=D-q$ and $\phi_i$ are scalar functions on the $p$-dimensional factor, the action (\ref{eq:SD}) can be written as 
\begin{equation}
S = \frac{1}{2\kappa^2_q}\int \mathrm{d}^q x \sqrt{-g_q} R[g_q]+\ldots \qquad \mathrm{where} \qquad
\frac{1}{2\kappa^2_q} \equiv \frac{1}{2\kappa^2_D}\int e^{(q-2)\phi_1+p\,\phi_2}\, \mathrm{vol}_p.
\end{equation}

When the $D$ dimensional metric is of the form
\begin{equation}\label{eq:Gred}
\mathrm{d}s_D^2 = e^{2\phi_1}\mathrm{d}s^2(\mathrm{AdS}_q) + e^{2\phi_2}\mathrm{d}s^2_p
\end{equation}
with $\dd s_p^2$ a metric on a compact space, this formula computes a dimension dependent ``holographic central charge'' $\mathcal{C}$, proportional to the dimensionless quantity
\begin{equation}\label{eq:cfn}
\mathcal{C} \propto \frac{L^{(q-2)}}{ G_N^{(q)}} \sim  \frac{8\pi}{\kappa_D^2}\int e^{(q-2)\phi_1+p\,\phi_2}\, \mathrm{vol}_p
\end{equation}
where $L$ is the AdS$_q$ radius and we have introduced the Newton constant $8\pi G_N^{(q)} = \kappa^2_q$. For example, when $q=3,4$ and 5 the holographic central charge $\mathcal{C}$ is proportional to the dual theory's anomaly coefficient $c$, three-sphere free energy $F_{S^3}$, and the anomaly coefficient $a$, respectively. These quantities are known to decrease monotonically along spatially homogenous RG flows in these theories.

If instead the metric (\ref{eq:Gred}) is asymptotically AdS$_D$ and holographically dual to a codimension two conformal defect, $q=D-2$, $p=2$ and the metric $\dd s^2_p$ is non-compact. Comparing to the ansatz (\ref{eq:gAnz}), we can formally make the identification $\phi_1 = V$, $\phi_2 = 0$, and $q=d-1$, and write
\begin{equation}
\mathcal{C} \sim \frac{2\pi}{G_N^{(d+1)}}\int \dd y\, e^{(d-3)V}\,|f h|.
\end{equation}
This is precisely the integral which appears\footnote{This integral also arises in application of the ``holographic central charge'' formula of \cite{Klebanov:2007ws,Macpherson:2014eza} for related reasons.} in (\ref{eq:Ceq}). It is thus tempting to conjecture, by analogy, that the appropriately defined ``universal part'' of the defect entanglement entropy plays a role similar to the holographic central charges when defects are present. In particular, that this quantity decreases along a spatially homogeneous deformation of the defect conformal field theory.

By comparing the value of $\mathcal{C}_\mathcal{D}$ in the putative UV and IR of such defect RG flows in three and four dimensions, we demonstrate that this not the case. Indeed, across the allowed parameter space of superconformal monodromy defects we consider, the quantity $\mathcal{C}_\mathcal{D}$ is shown to both increase as well as decrease in the IR (and may even vanish)\footnote{In fact, even for RG flows localised on the defect it is known that this quantity need not be monotonic e.g. \cite{Kobayashi:2018lil,Casini:2023kyj,Jokela:2025qac}. }. Nevertheless, as we show in appendix \ref{app:RG} there is a sense in which a scheme-independent part of the sphere entanglement entropy in the dCFT ``at large'' does in fact decrease from UV to IR in these flows.

There are various interesting extensions of these calculations that are motivated by our work. The conformal defects we investigate here are known to admit generalizations with superconformal spatially varying mass deformations by operators in the ambient theory.  These defect conformal field theories are studied holographically in \cite{Arav:2024wyg,Arav:2024exg}. While the dual gravitational solutions are only known numerically, the quantities $h_D$, $\mathcal{I}_D$ and $b$ are known analytically as functions of the defect parameters $n$ and $g\mu^I$. It would be interesting to verify\footnote{Preliminary investigation suggests that this is indeed the case. We thank Yusheng Jiao for communicating his results with us.} whether or not the defect entanglement entropy in these more complicated defect theories also satisfies (\ref{eq:dS4d}), (\ref{eq:dS5d}).

Even within the relatively simple class of holographic monodromy defects catalogued in appendix \ref{app:sols} there remains room for further exploration. In particular, we have focused entirely on monodromy defects which preserve some supersymmetry. As explained in the appendix, however, the solutions reviewed in sections \ref{app:4dSTU}, \ref{app:5dSTU} and \ref{app:7dSTU} all admit simple supersymmetry breaking deformations. It would be interesting to work out in detail the defect theory observables for the non-supersymmetric holographically dual theories, especially in the context of the defect entanglement entropy.

The techniques employed in this work can also be brought to bear on the rich set of holographic monodromy defects obtained via double Wick rotation of known supersymmetric black holes in diverse dimensions. Examples along this line appear in \cite{Conti:2025}, where they are further explored from the perspective of their uplift to supergravity in ten dimensions. Somewhat orthogonally, it is also interesting to adapt these techniques to the study of monodromy defects in systems with reduced spacetime symmetries. This is the approach taken in \cite{Conti:2025x2}, where non-conformal defects are investigated holographically.

\section*{Acknowledgements}
We are indebted to Igal Arav, Jerome Gauntlett, Yusheng Jiao and Matthew Roberts for collaboration on related topics and enlightening discussion. AC and YL are partially supported by the grants from the Spanish government MCIU-22-PID2021-123021NB-I00 and MCIU-25-PID2024-161500NB-I00. The work of AC is also supported by the Severo Ochoa fellowship PA-23-BP22-019, as well as the INGENIUM Alliance of European Universities which enabled continued training at the University of Crete. AC thanks the Crete Center for Theoretical Physics and Imperial College London for the kind hospitality while some parts of this work were being prepared.  The work of CR is supported through the framework of H.F.R.I. call ``Basic research Financing (Horizontal support of all Sciences)'' under the National Recovery and Resilience Plan ``Greece 2.0'' funded by the European Union--NextGenerationEU (H.F.R.I. Project Number: 15384). Y.L. would like to thank the Isaac Newton Institute for Mathematical Sciences, Cambridge, for support and hospitality during the programme "Boundaries, Impurities and Defects" where work on this paper was undertaken. This work was supported by EPSRC grant no EP/R014604/1.$\& \#$34

\begin{appendix}
\section{Holographic monodromy defects and supergravity}\label{app:sols}
The monodromy defects investigated in this work all have holographic duals in solutions to simple truncations of maximal gauged supergravities. The bosonic sector of these truncations retains the gauge fields in the Cartan of the gauge group. From the perspective of $\mathcal{N}=2$ supersymmetry, one of these vectors resides in the graviton multiplet while the others fill out vector multiplets. Such truncations are sometimes referred to as ``STU'' or ``STU-like'' theories.

In this appendix we catalogue the general holographic monodromy defect solutions studied in this work, in various dimensions. These solutions can be obtained by double Wick rotation of static black holes in $D=4, 5$, and 7 gauged supergravity \cite{Cvetic:1999xp,Lu:2003iv,Cvetic:1999ne}. Our conventions are mostly aligned with those of \cite{Ferrero:2021etw}.
\subsection{Four dimensions}\label{app:4dSTU}
The general $D=4$ STU defect solution with arbitrary monodromy sources can be written \cite{Ferrero:2021etw}
\begin{equation}
\begin{split}\label{eq:4dU14solution}
\dd s^2_4 &  = \, \frac{H^{1/2}}{4 g^2} \Big[ \frac{1}{4} \dd s^2(\text{AdS}_2) +\frac{1}{P} \dd y^2+\frac{P}{4H} n^2 \dd z^2 \Big], \\[2mm]
A^I & = \left( \alpha^I - \sqrt{1 - \frac{\nu}{q_I}}\frac{n}{2 g} \frac{y}{h_I} \right) \dd z\,, \qquad
X^I  = \, \left(\frac{H}{h_I^4}\right)^{1/4}.
\end{split}
\end{equation}
Here $ \dd s^2(\text{AdS}_2) $ is a unit radius metric on AdS$_2$, and $h_I$, $H$ and $P$ are functions that depend only on $y$. They are given by
\begin{equation}\label{eq:functions4DU14}
h_I = y+q_I , \qquad H = h_0 h_1 h_2 h_3 , \qquad P = H - 4 y^2 - 4 \nu y.
\end{equation}
The $q_I$ are four constants and all the $h_I$ must have the same sign, either positive or negative. The parameter $\nu$ breaks supersymmetry---in other words, if we set $\nu = 0$ the solution preserves two Poincar\'e and two superconformal supersymmetries. Note that if one sets $q_I = 0$ and $n = 1$ we obtain the maximally symmetric AdS$_4$ vacuum with radius $L = 1/2 g$. 

In our conventions we take
\begin{equation}
\alpha^I = \mu_I + \frac{n}{2g}
\end{equation}
and highlight two convenient bases for the monodromy parameters. In the first, we have
\begin{align}\label{eq:mu4d}
g\mu_R & = (1-n) = g\left(\mu_0 + \mu_1 + \mu_2 + \mu_3\right) & g\mu_{F_1} = g\left(\mu_1-\mu_2 \right)\nonumber\\
g\mu_{F_2} & = g\left( \mu_2-\mu_3\right) & g\mu_{F'} = \frac{1}{2}g\left(3\mu_0-\mu_1-\mu_2-\mu_3 \right)
\end{align}
while in the second, which is relevant for putative flows to an mABJM monodromy defect is
\begin{align}
g\mu_R & = (1-n) = g\left(\mu_0 + \mu_1 + \mu_2 + \mu_3\right) & g\mu_{F_1} = g\left(\mu_1-\mu_2 \right)\nonumber\\
g\mu_{F_2} & = g\left( \mu_2-\mu_3\right) & g\mu_{B} = g\left(\mu_0-\mu_1-\mu_2-\mu_3 \right).
\end{align}
The relationship between the monodromy source for the $R$-symmetry $\mu_R$ and the conical singularity parameter $n$ is imposed by the preserved supersymmetry.

This STU model admits various simple subtruncations. In particular, setting the gauge fields equal to one another ($A^0=A^1=A^2=A^3$) and the scalars to zero ($X^I = 1$) results in a truncation to minimal gauged supergravity. The holographic monodromy defect solution in this subtruncation can be obtained from (\ref{eq:4dU14solution}) by taking
\begin{equation}\label{eq:functions4Dmin}
h_I = y+q. 
\end{equation}
The minimal gauged supergravity defect has a non-vanishing monodromy source $\mu_R$ only when the dual dCFT has a conical singularity at the location of the defect (i.e. when $n \ne 1$).

\subsection{Five dimensions}\label{app:5dSTU}
The general $D=5$ STU defect solution with arbitrary monodromy sources  can be written \cite{Kunduri:2007qy,Ferrero:2021etw}
\begin{equation}\label{eq:5DU13solution}
\begin{split}
\dd s^2_5 & = \frac{H^{1/3}}{g^2} \left[\dd s^2(\text{AdS}_3)+\frac{1}{4P} \dd y^2+\frac{P}{H} n^2 \dd z^2\right], \\[2mm]
A^I & =  \left( \alpha^I - \sqrt{1- \frac{\nu}{q_I}} \frac{n}{g}\frac{y}{h_I}\right) \dd z , \qquad X^I = \frac{H^{1/3}}{h_I}.
\end{split}
\end{equation}
Here $\dd s^2(\text{AdS}_3)$ is a unit radius metric on AdS$_3$, and $h_I, H$ and $P$ are functions that depend only on $y$. They are given by
\begin{equation}\label{eq:5DU13functions}
h_I = y + q_I  ,\qquad H =  h_1 h_2 h_3 , \qquad P = H - y^2 - \nu y,
\end{equation}
where $q_I$ are constants. Note that we recover the maximally symmetric AdS$_5$ vacuum solution with  $L=1/g$ if we set $q_I = 0$ and $n=1$. When the parameter $\nu$ vanishes, the solutions generically preserve $\mathcal{N}=(0,2)$ superconformal symmetry in two dimensions.

In our conventions, we take
\begin{equation}
\alpha^I = \mu_I +\frac{n}{g}
\end{equation}
and there are two useful bases for the monodromy sources that are referred to in this work. In the first, we define
\begin{equation}
g\mu_R = (1-n) = g\left(\mu_1+\mu_2 + \mu_3 \right), \quad g\mu_F = g\left(\mu_1 -\mu_2 \right), \quad g\mu_{F'} = \frac{2}{3}g\left(\mu_1+\mu_2-2\mu_3 \right)
\end{equation}
In the second, which is relevant for putative RG flows to an LS monodromy defect, we have instead
\begin{equation}
g\mu_R = (1-n) = g\left(\mu_1+\mu_2 + \mu_3 \right), \quad g\mu_F = g\left(\mu_1 -\mu_2 \right), \quad g\mu_{B} = g\left(\mu_1+\mu_2-\mu_3 \right).
\end{equation}
The relationship between the monodromy source for the $R$-symmetry $\mu_R$ and the conical singularity parameter $n$ is imposed by the preserved supersymmetry.

Again, this supergravity theory admits various simple subtruncations. When the gauge fields are all set equal to one another ($A^0=A^1=A^2$) and the scalars vanish ($X^I = 1$), one obtains minimal gauged supergravity in five dimensions. The holographic monodromy defect solution in this truncation can be obtained from (\ref{eq:5DU13solution}) by taking
\begin{equation}\label{eq:functions5Dmin}
h_I = y+ q.
\end{equation}
The minimal gauged supergravity defect again has a non-vanishing monodromy source $\mu_R$ only when the dual dCFT has a conical singularity at the location of the defect.

\subsection{Seven dimensions}\label{app:7dSTU}
The general $D=7$ STU defect solution with arbitrary monodromy sources can be written \cite{Ferrero:2021wvk,Gutperle:2022pgw}
\begin{equation} \label{eq:D7U12solution}
\begin{split}
\dd s^2_7 & = \frac{(yH)^{1/5}}{g^2} \left[\dd s^2(\text{AdS}_5) +\frac{y}{P}\dd y^2+\frac{P}{H} n^2 \dd z^2\right] , \\[2mm]
A^I & = \left(\alpha^I - \sqrt{1-\frac{\nu}{q_I}}\frac{2}{g} n \frac{y^2}{h_I} \right) \dd z , \qquad X^I = \frac{(yH)^{2/5}}{h_I}.
\end{split}
\end{equation}
Here $\dd s^2(\text{AdS}_5)$ is a unit radius metric on AdS$_5$, and $h_I, H$ and $P$ are functions that depend only on $y$. They are given by
\begin{equation}
h_I = y^2+q_I ,\qquad H  = h_1 h_2 , \qquad P = H-4 y^3 - 4 \nu y,
\end{equation}
where $q_I$ are constants. Note that we recover the maximally symmetric AdS$_7$ vacuum solution with  $L=2/g$ if we set $q_I = 0$ and $n=1$. When the parameter $\nu$ vanishes, the solutions generically preserve $\mathcal{N}=1$ superconformal symmetry in four dimensions.

In our conventions, we take
\begin{equation}
\alpha^I = \mu_I + \frac{2n}{g}
\end{equation}
as well as
\begin{equation}
g\mu_R = (n-1) = -g(\mu_1+\mu_2) \qquad \mathrm{and} \qquad g\mu_F = -g(\mu_2-\mu_1).
\end{equation}
Once again, the relationship between the monodromy source for the $R$-symmetry $\mu_R$ and the conical singularity parameter $n$ is imposed by the preserved supersymmetry.

\section{Holographic regularization and limits of integration}\label{app:cut}
As emphasised in the main text, the computation of the defect entanglement entropy $\mathcal{C}_{\mathcal{D}}^{(d)}$ for the superconformal monodromy defects studied in this work revolves around evaluation of the expression (\ref{eq:Ceq}). The crux of this computation is a proper handling of the divergences encountered in the integral expression
\begin{equation}\label{eq:divI}
\left[ \int \dd y\, e^{(d-3)V} | f h |\right]_\mathcal{D}.
\end{equation}
In this appendix we briefly review the prescription of \cite{Jensen:2013lxa} for regulating these divergences, and provide various intermediate steps useful in obtaining our main results.

The holographic monodromy defects surveyed in appendix \ref{app:sols} share several broadstroke features. Namely, there is a location in the bulk, at $y=y_c$, in which the solution smoothly terminates. We will refer to this location as  the ``core'', and it is regular in the following sense:

The metric on these spacetimes in the $y$--$z$ directions can generically be written
\begin{equation}
\dd s^2\Big|_{(y,z)} = e^{2\Omega}\left[\frac{\dd y^2}{P}+C \frac{P}{H}n^2\dd z^2 \right]
\end{equation}
where $\Omega$ and $C$ are functions of $y$. The metric function $P$ is a function of $y$ which has an outermost root (the largest) at $y=y_c$. In the neighbourhood of this root, the various functions generically behave to leading order like
\begin{equation}
P = P_1(y-y_c)+\ldots \qquad \Omega = \Omega_0 +\ldots \qquad C = C_0 +\ldots\qquad H = H_0 + \ldots
\end{equation}
and thus, introducing the coordinate $\rho$ such that $\dd \rho = \dd y /\sqrt{P}$, one finds that near the solution's core this part of the metric behaves like
\begin{equation}
\dd s^2\Big|_{(y,z)} = e^{2\Omega_0}\left[\dd \rho^2+C_0 \frac{P_1^2}{4H_0}n^2\,\rho^2\dd z^2 \right].
\end{equation}
Keeping in mind the periodicity of $z$, $\Delta z = 2\pi$, we then find that a regular core implies the constraint
\begin{equation}
n^2 = \frac{4H_0}{C_0 P_1^2}.
\end{equation}
Regularity of the gauge fields then amounts to the requirement that
\begin{equation}
A^I(y_c) = 0.
\end{equation}

The domain of the $y$ coordinate is $y\in [y_c,\infty]$, and it is straightforward to see that for the solutions of appendix $\ref{app:sols}$ it is the upper limit of this domain which leads to divergences in the individual terms in the expression given in (\ref{eq:divI}). To regulate these divergences, we find it convenient to introduce a holographic cut-off at $y=y_\epsilon$. 

To do so, we first bring the asymptotic region of the metric (near the AdS$_{d+1}$ boundary) to a canonical Fefferman-Graham (FG) form,
\begin{multline}
\mathrm{d}s^2_{d+1} = L^2\frac{\mathrm{d}\zeta^2}{\zeta^2}
+\frac{L^2}{\zeta^2}\Bigg[\mathfrak{g}_1\left(\frac{\zeta}{r_\perp}\right)\left(-\dd t^2+\dd r_{||}^2 +r_{||}^2\,\dd \Omega_{d-4}^2\right)\\
+ \mathfrak{g}_2\left(\frac{\zeta}{r_\perp}\right)\mathrm{d}r_\perp^2+\mathfrak{g}_3\left(\frac{\zeta}{r_\perp}\right)r_\perp^2n^2\mathrm{d}z^2\Bigg]
\end{multline}
where the functions, to leading order, behave like
\begin{equation}
\mathfrak{g}_i = 1+\mathcal{O}\left(\frac{\zeta}{r_\perp}\right)
\end{equation}
and we have allowed for the presence of a conical defect on the boundary parametrised by the constant $n$.

The coordinate transformation which accomplishes this manoeuvre is reported in \cite{Jensen:2013lxa} to be given by
\begin{equation}\label{eq:FG1}
\zeta = Z\,\mathfrak{G}(y), \qquad r_\perp = Z\, \mathfrak{F}(y)
\end{equation}
where
\begin{equation}
\mathfrak{G} = e^{-\frac{1}{L}\int^y\mathrm{d}y'f\sqrt{1-L^2e^{-2V}}}\qquad \mathrm{and} \qquad \mathfrak{F} = e^{-L\int^y\mathrm{d}y'f\frac{e^{-V}}{\sqrt{e^{2V}-L^2}}}.
\end{equation}
It is easy to demonstrate that this transformation has the desired effect. 

Once this transformation has been performed, the standard prescription is simple to state: a cut-off is implemented on the FG radial coordinate $\zeta = \epsilon$ and the coordinate transformation can then be inverted (at non-vanishing $Z$) to find the corresponding location in the original coordinate $y_\epsilon$.

An important subtlety is that the change in coordinates (\ref{eq:FG1}), when used in the context of the entanglement entropy calculation (\ref{eq:S_EEd4p}), is to be evaluated on the minimal surface obeying $Z^2 + r_{||}^2 = R^2$. It is thus evident that the cut-off $y_\epsilon$ is generically a function of two arguments, $y_\epsilon(r_{||},R)$. Notably, the $r_{||}$ dependence of the cut-off impedes the factorisation of the two integrals in  (\ref{eq:S_EEd4p}).

By direct computation, however, we find that for the holographic monodromy defects studied here\footnote{Indeed we expect that this result can be proven to hold with more generality.} the dependence of (\ref{eq:divI}) on $r_{||}$ completely drops out in the vacuum subtraction. Thus, in this context, the $y$ and $r_{||}$ integrals factorize under the action of $[\ldots]_\mathcal{D}$, the latter simply giving rise to a dimension dependent prefactor governed by the boundary entanglement geometry.

We now focus our attention on the specific solutions of Appendix \ref{app:sols}, providing the limits of integration for the evaluation of (\ref{eq:divI}) using the prescription outlined above.
\subsection{Four dimensions}
In the coordinates of (\ref{eq:4dU14solution}), the UV cut-off can be written
\begin{equation}
y_\epsilon = \frac{Z}{\epsilon}-\frac{1}{4}\left(q_1+q_2+q_3+q_4 \right)-\frac{1}{32}\left(q_1+q_2+q_3+q_4 \right)^2\frac{\epsilon}{Z} + \ldots
\end{equation}
where $Z = R$ on the entangling surface.

The core value, written in terms of the conical singularity parameter and flavor monodromies, is given by
\begin{multline}
y_{c}  = \frac{1}{6\sqrt{3} n^2} \sqrt{2 g \mu_{F'}+ n+1} \sqrt{2 g (4 \mu_{F_1} + 2 \mu_{F_2} - \mu_{F'} ) + 3 (n+1)}\,\times \\ \sqrt{3 (n+1)-2 g (2 \mu_{F_1} - 2 \mu_{F_2} + \mu_{F'})} \sqrt{3(n+1)-2 g (2 \mu_{F_1} + 4 \mu_{F_2} + \mu_{F'})},
\end{multline}
and the $q_I$ can similarly be written as
\begin{align}
q_1 & = \frac{(n-1-2 g \mu_{F'})}{6 \sqrt{3} n^2 \sqrt{1 + n +2 g \mu_{F'}}} \sqrt{2 g (4 \mu_{F_1} + 2 \mu_{F_2} - \mu_{F'} )+3 (n+1)}\, \times \nonumber\\
&\sqrt{3 (n+1)-2 g (2 \mu_{F_1} - 2 \mu_{F_2} + \mu_{F'})}\sqrt{3 (n+1)-2 g (2 \mu_{F_1} + 4 \mu_{F_2} + \mu_{F'})}\nonumber \\
q_2 & =  \frac{(-2 g (4 \mu_{F_1} + 2 \mu_{F_2} - \mu_{F'}) - 3 (1-n))}{6 \sqrt{3} n^2 \sqrt{2 g (4 \mu_{F_1} + 2 \mu_{F_2} - \mu_{F_p} ) + 3 (n+1)}} \sqrt{2 g \mu_{F'} + n+1} \, \times \nonumber\\
&\sqrt{3 (n+1)-2 g (2 \mu_{F_1} - 2 \mu_{F_2} + \mu_{F'})} \sqrt{3 (n+1)-2 g (2 \mu_{F_1} + 4 \mu_{F_2} + \mu_{F'})} \nonumber\\
q_3 & = \frac{(2 g (2 \mu_{F_1} - 2 \mu_{F_2} + \mu_{F'})-3 (1-n))}{6 \sqrt{3} n^2 \sqrt{3 (n+1)-2 g (2 \mu_{F_1} - 2 \mu_{F_2} + \mu_{F'})}} \sqrt{2 g \mu_{F'}+n+1}\, \times   \nonumber \\
& \sqrt{2 g (4 \mu_{F_1} + 2 \mu_{F_2} - \mu_{F'})+3 (n+1)}\sqrt{3 (n+1)-2 g (2  \mu_{F_1} + 4 \mu_{F_2} + \mu_{F_p})} \nonumber\\
q_4 & = \frac{(2 g (2 \mu_{F_1}+4 \mu_{F_2} + \mu_{F'})-3 (1-n))}{6 \sqrt{3} n^2 \sqrt{3 (n+1)-2 g (2 \mu_{F_1}+4 \mu_{F_2} + \mu_{F'})}} \sqrt{2 g \mu_{F'}+n+1}\, \times \nonumber\\
&  \sqrt{2 g ( 4 \mu_{F_1} + 2 \mu_{F_2} - \mu_{F'}) +3 (n+1)} \sqrt{3 (n+1)-2 g (2 \mu_{F_1} - 2 \mu_{F_2} + \mu_{F'})}.
\end{align}

\subsection{Five dimensions}
In the coordinates of (\ref{eq:5DU13solution}), the UV cut-off can be written
\begin{equation}
y_\epsilon = \frac{Z^2}{\epsilon^2}-\frac{1}{3}\left(q_1+q_2+q_3 \right)-\frac{1}{18}\left(q_1+q_2+q_3 \right)^2\frac{\epsilon^2}{Z^2}+\ldots
\end{equation}
where $Z = \sqrt{R^2-r_{||}^2}$ on the entangling surface.

The core value, written in terms of the conical singularity parameter and flavor monodromies, is given by
\begin{equation}
y_{c} = \frac{( 2 ( 1 +2 n)-3 g \mu_{F'}) (3 g (\mu_{F'} - 2 \mu_{F} ) +4 (1+ 2n)) (3 g (2 \mu_{F} + \mu_{F'} )+ 4 (1+ 2n))}{864 n^3},
\end{equation}
and the $q_I$ can similarly be written as
\begin{align}
q_1 & = \frac{(2 (1 + 2n) -3 g \mu_{F'} ) (3 g (\mu_{F'} - 2 \mu_{F})+4 (1 + 2n) (-3 g (2 \mu_{F} + \mu_{F'})+4 (n-1))}{864 n^3}\nonumber\\
q_2 & = \frac{( 2 (1 + 2n) -3 g \mu_{F'} ) (3 g ( 2\mu_{F}- \mu_{F'}) +4 (n-1) ) (3 g (2 \mu_{F} + \mu_{F'})+ 4( 1+ 2n))}{864 n^3} \nonumber \\
q_3 & = \frac{(3 g \mu_{F'} + 2 (n-1) ) (3 g (\mu_{F'} - 2 \mu_F)+ 4 (1 + 2n)) (3 g (2 \mu_{F} + \mu_{F'})+4 (1 + 2n))}{864 n^3}.
\end{align}
\subsection{Seven dimensions}

In the coordinates of (\ref{eq:D7U12solution}), the UV cut-off can be written
\begin{equation}
y_\epsilon = \frac{Z^2}{\epsilon^2}-\frac{1}{5}\left(q_1 + q_2 \right)\frac{\epsilon^2}{Z^2}-\frac{3}{50}\left(q_1 + q_2 \right)^2\frac{\epsilon^6}{Z^6}+\ldots
\end{equation}
where $Z = \sqrt{R^2-r_{||}^2}$ on the entangling surface.

The core value, written in terms of the conical singularity parameter and flavor monodromy source, is given by
\begin{equation}\label{eq:yc7d}
y_{c}  = \frac{(1 + 3 n)^2-g^2 \mu_F^2}{4 n^2} ,
\end{equation}
and the $q_I$ can be written
\begin{align}
q_1 & = \frac{(n-1-g \mu_{F}) \left((1 + 3 n)^2-g^2 \mu_{F}^2 \right)^2}{16 n^4 (g \mu _F + 1 + 3n )} \nonumber\\
q_2 & = \frac{(1- n - g \mu_F) \left((1 + 3 n)^2-g^2 \mu_F^2\right)^2}{16 n^4 (g \mu_F -3 n-1)}.
\end{align}

\subsubsection{Holographic renormalization, one-point functions and the on-shell action}\label{sec:appOS}
In order to compute various one-point functions for the holographic defects in the 6$d$ $\mathcal{N}=(2,0)$ theory, we employ the machinery of holographic renormalization \cite{deHaro:2000vlm, Papadimitriou:2016yit}. We will perform this calculation for the theory on $AdS_5\times S^1$---the results for an $\mathbb{R}^{1,5}$ background follow via Weyl transformation. This analysis overlaps with the computation of the stress tensor one-point function and on-shell action in \cite{Gutperle:2022pgw,Capuozzo:2023fll}, though the perspective taken here is slightly different.

In our conventions, the gravitational action which gives rise to the equations of motion solved by (\ref{eq:D7U12solution}) can be written
\begin{equation}
S = S_B + S_{GH} + S_{CT}+S_{CT'}.
\end{equation}
Here, the bulk action and Gibbons-Hawking-York term (integrated over the holographic boundary) are given by
\begin{multline}
S_B+S_{GH} = \frac{1}{16\pi G_N^{(7)}}\int \dd^7x\sqrt{-g}\left(R-g^2\mathcal{V}-\frac{1}{2}\partial\vec{\phi}\cdot\partial\vec{\phi} -\frac{1}{4}X_I^{-2} F_I^2\right)\\+\frac{1}{8\pi G_N^{(7)}}\int \dd^6 x\sqrt{-\gamma} K,
\end{multline}
where
\begin{equation}
X_I = e^{-\frac{1}{2}\vec{a}_I\cdot\vec{\phi}} \qquad \mathrm{with}\qquad \vec{a}_1 = \left(\sqrt{2},\sqrt{\frac{2}{5}}\,\right), \qquad \vec{a}_2 = \left(-\sqrt{2},\sqrt{\frac{2}{5}}\,\right)
\end{equation}
and $F_I = \dd A^I$. Note that $I$ indices are raised and lowered with the identity. The scalar potential $\mathcal{V}$ can be written in terms of a superpotential $\mathcal{W}$ like
\begin{equation}
g^2\mathcal{V} = 2\,\delta^{IJ}\partial_I\mathcal{W}\partial_J\mathcal{W}-\frac{6}{5}\mathcal{W}^2
\end{equation}
with
\begin{equation}
\mathcal{W} = -g\left(X_1+X_2+\frac{1}{2}\frac{1}{X_1^2 X_2^2} \right).
\end{equation}
Additional boundary terms are needed to holographically renormalize the bulk action. These include the divergent counterterms
\begin{equation}
S_{CT} = \frac{1}{16\pi G_N^{(7)}}\int \dd^6 x\sqrt{-\gamma}\left(2\mathcal{W} -\frac{1}{2g}R[\gamma]-\frac{1}{4g^3}\left(R[\gamma]_{ij}R[\gamma]^{ij}-\frac{3}{10}R[\gamma]^2 \right)+\ldots\right)
\end{equation}
where the ellipses contain terms that play no role in the present analysis, as well as finite counterterms 
\begin{equation}
S_{CT'} = \frac{1}{16\pi G_N^{(7)}}\int \dd^6 x\sqrt{-\gamma}\left(\frac{1}{160 g^5}\Big(\delta_B R[\gamma]R[\gamma]_{ij}R[\gamma]^{ij}+\delta_C R[\gamma]^3 \Big)+\ldots \right).
\end{equation}
Again, these are not the full set of possible finite counterterms\footnote{For a basis of the purely gravitational 6 derivative terms, see e.g. \cite{Bonora:2023yza}. We leave a more comprehensive study of the allowed finite counterterms and their interplay with supersymmetry to future work.}, but as a consequence of the isometries enjoyed by the solutions we study here, these terms provide a convenient proxy.

Under this holographic renormalisation scheme, we eventually arrive at the following one-point functions for the holographic defect theories dual to (\ref{eq:D7U12solution}):

\begin{align}
\langle J^R \rangle  = -\frac{1}{2}\left(\langle J^1 \rangle+\langle J^2 \rangle \right) & =\frac{n}{g^5 8\pi G_N^{(7)}} (q_1+q_2) \label{eq:JR7}\\
\langle J^F \rangle  = -\frac{1}{2}\left(\langle J^2 \rangle-\langle J^1 \rangle \right) & =\frac{n}{g^5 8\pi G_N^{(7)}} (q_2-q_1)\label{eq:JF7}\\
\langle \mathcal{T}_{ij} \rangle\,\dd x^i\dd x^j & = -\frac{4(q_1+q_2)-5(2+\delta_B+5\delta_C)}{40g^5\pi G_N^{(7)}}\left[\dd s^2(AdS_5)-5 n^2 \dd z^2 \right]
\end{align}
To fix the coefficients of the finite counterterms, $\delta_{B,C}$, we appeal to the fact that in three and four dimensions superconformal monodromy defects preserving at least two Poincar\'e supercharges are characterized by a conformal weight proportional to the $R$-symmetry current \cite{Bianchi:2019sxz,Arav:2024wyg}. Assuming this is the case also in six dimensions, this can be arranged provided that 
\begin{equation}\label{eq:susyRN}
\delta_B+5\,\delta_C = -2
\end{equation}
in which case 
\begin{equation}
\langle \mathcal{T}_{ij}\, \rangle\dd x^i \dd x^j = -\frac{h_D}{2\pi}\left[\dd s^2(AdS_5)-5 n^2 \dd z^2 \right]
\end{equation}
with
\begin{equation}
h_D = \frac{L^5}{160 G_N^{(7)}}\left( q_1 + q_2\right).
\end{equation}

Finally, we compute the value of the on-shell action using our holographic renormalization scheme. Towards this end, it is expedient to use the fact that the isometries of  (\ref{eq:D7U12solution}) allow one to express $S_B$ as a total derivative on-shell\footnote{See e.g. footnote 21 of \cite{Arav:2024wyg}.}. In particular, we find
\begin{equation}
S_B^{os} = \frac{1}{16\pi G_N^{(7)}}\int \dd^7 x\sqrt{-g}\,\nabla_\nu q^{z\nu} \qquad \mathrm{where}\qquad q_{\mu\nu} = 2\nabla_\mu k_\nu + \sum_I X_I^{-2} A^I_\rho k^\rho F^I_{\mu\nu}
\end{equation}
and $k = \partial_z$ a Killing vector of the background solution.

For metrics of the form (\ref{eq:gAnz}) and writing the gauge fields as $A^I = a_I \dd z$ with the $a_I$ functions of $y$, one obtains
\begin{equation}
S_B^{os} = -\frac{1}{8 G_N^{(7)}}\int\star_5\,e^{5V}\left[\frac{2h'}{f}+\frac{1}{fh}\left(\frac{a_1'a_1}{X_1^2}+\frac{a_2'a_2}{X_2^2}\right) \right]\Bigg|^{y_\epsilon}_{y_c},
\end{equation}
where $\star_5$ is the Hodge dual taken with respect to the unit radius AdS$_5$. 

Evaluated and summed together with $S_{GH}$, $S_{CT}$ and $S_{CT'}$ on-shell, one obtains 
\begin{multline}
S^{os} = \frac{1}{4g^5 G_N^{(7)}}\mathrm{vol}_5\Big[2g(\alpha_1 q_1+\alpha_2 q_2)-5n(2+\delta_B+5\delta_C) \Big]+\frac{1}{8G_N^{(7)}}\mathrm{vol}_5 \,e^{5V}\frac{2h'}{f}\Bigg|^{y_c}\\
= \frac{1}{4g^5 G_N^{(7)}}\mathrm{vol}_5\Bigg[2g(\alpha_1 q_1+\alpha_2 q_2)-5n(2+\delta_B+5\delta_C)+2 y_c^2 \Bigg]
\end{multline}
where $\mathrm{vol}_5$ is the (appropriately regulated) volume of a unit radius AdS$_5$. Using the renormalisation scheme (\ref{eq:susyRN}), this can be written
\begin{align}
S^{os} &= \frac{1}{2g^5 G_N^{(7)}}\mathrm{vol}_5\Bigg[g(\alpha_1 q_1+\alpha_2 q_2)+ y_c^2 \Bigg]\\
& = 2\pi\,\mathrm{vol}_5\frac{1}{n}\Bigg[4n\langle J^R \rangle-g\mu_R \langle J^R \rangle-g\mu_F\langle J^F\rangle\Bigg]+2\pi\,\mathrm{vol}_5\frac{a^{(2,0)}}{96}y_c^2
\end{align}
where we have used the definitions
\begin{equation}
g\mu_R = (n-1) = -(\mu_1+\mu_2) \qquad \mathrm{and} \qquad g\mu_F = -(\mu_2-\mu_1)
\end{equation}
as well as (\ref{eq:JR7}), (\ref{eq:JF7}) and (\ref{eq:a6d}). Further substituting (\ref{eq:yc7d}) into this expression allows one to write the renormalized on-shell action entirely in terms of field theory quantities. In fact, expressing the on-shell action entirely in terms of the conical singularity parameter and the flavor monodromy source yields
\begin{equation}
S^{os} = \mathrm{vol}_5 \frac{\pi}{48}a^{(2,0)}\, n \,y_c^2
\end{equation}

\section{Sphere entanglement entropy and defect RG flows}\label{app:RG}
In sections \ref{sec:RG4d} and \ref{sec:RG5d} we demonstrated that the defect contribution to the sphere entanglement entropy, $\mathcal{C}_\mathcal{D}$, does not necessarily decrease along an RG flow driven by spatially homogeneous deformations away from the defect. It is interesting to wonder whether or not the same is true for the universal part of the sphere entanglement entropy in these codimension two dCFTs ``at large''.

Towards this end, in this appendix we study the quantity $\mathcal{C}^{(d)}$ defined as
\begin{equation}
\mathcal{C}^{(d)} \equiv \mathcal{C}^{(d)}_\mathcal{D} + n\, \mathcal{C}^{(d)}_0
\end{equation}
where $\mathcal{C}^{(d)}_0$ is the universal part of the sphere entanglement entropy evaluated in the (defectless) CFT vacuum. In particular, we evaluate this quantity in both the UV and IR of the putative dCFT RG flows of sections \ref{sec:RG4d} and \ref{sec:RG5d}. We show that in both cases
\begin{equation}
\mathcal{C}^{\mathrm{dCFT}}_\mathrm{UV} >\mathcal{C}^{\mathrm{dCFT}}_\mathrm{IR}
\end{equation}
for all allowed values of the defect parameters $n$ and $g\mu^I$.
\subsection{dABJM RG flow}
This putative flow is driven from the dABJM theory by a spatially homogeneous mass deformation. It is conjectured to arrive at the defect mABJM theory in IR.

\begin{figure}[h]
\centering
\includegraphics[scale=0.6]{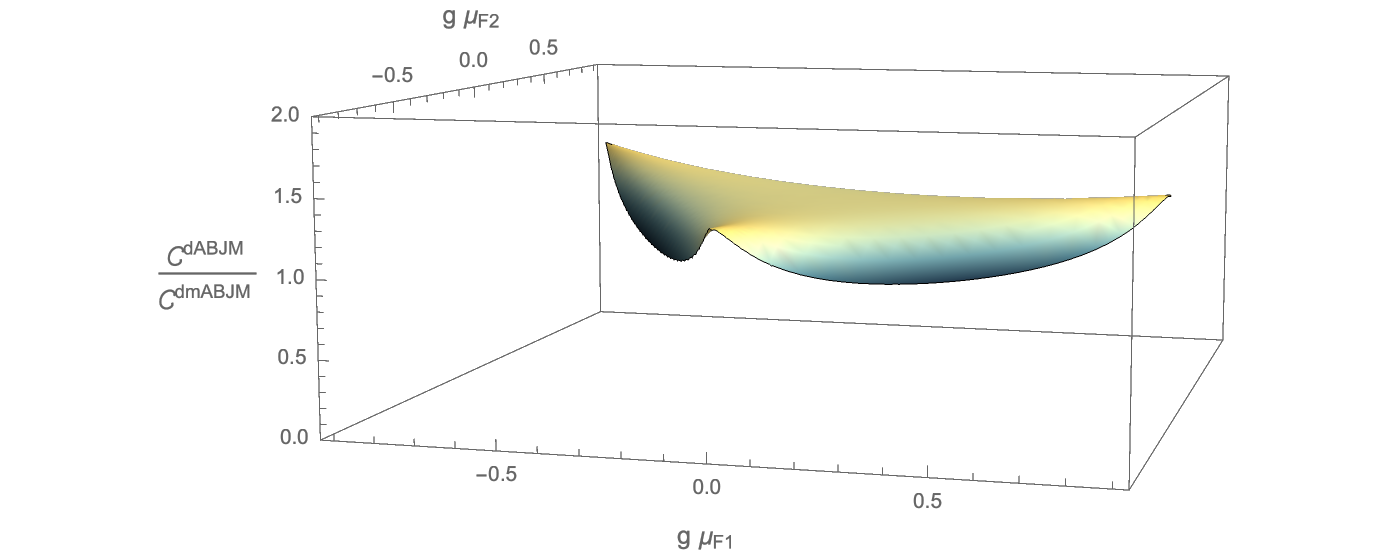}
\caption{The ratio of $\mathcal{C}^{(3)}$ between ABJM and mABJM (with $\mu_B=0$)  defects for $n=1$. This quantity is greater than one for all allowed values of the monodromy sources.}\label{fig:Cd4Dratnis1}
\end{figure}

\begin{figure}[h]
\centering
\includegraphics[scale=0.65]{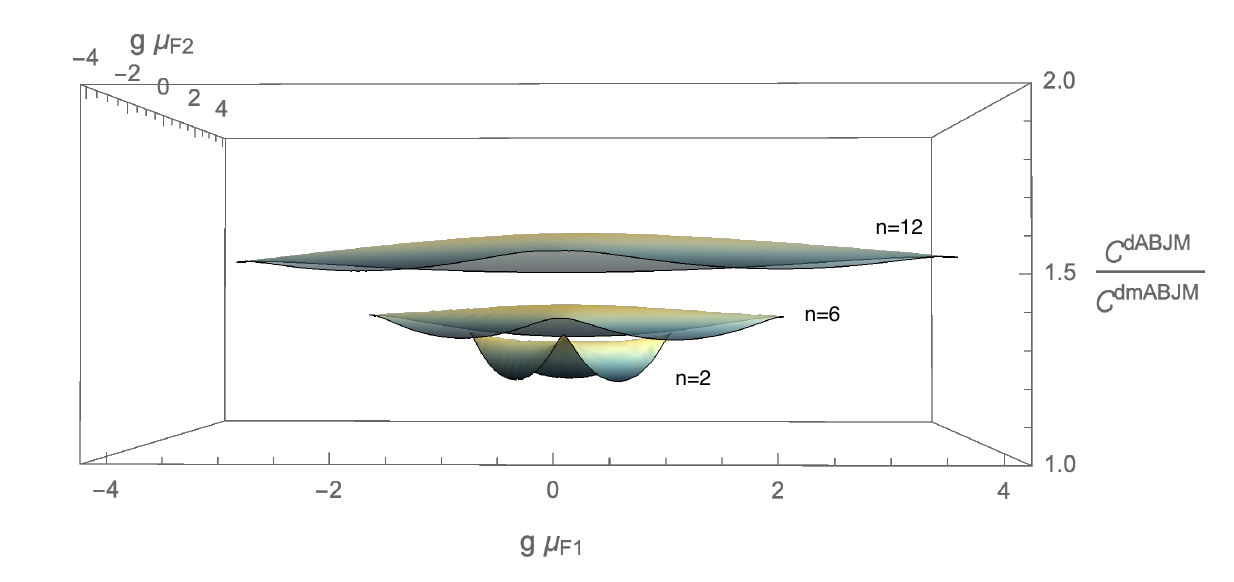}
\caption{The ratio of $\mathcal{C}^{(3)}$ between ABJM and mABJM (with $\mu_B=0$)  defects for various values of $n>1$. This quantity is greater than one for all allowed values of the monodromy sources.}\label{fig:Cd4Drat}
\end{figure}

\subsection{d$\mathcal{N}=4$ SYM RG flow}
This putative flow is driven from the defect $\mathcal{N}=4$ SYM theory by a spatially homogeneous mass deformation. It is conjectured to arrive at the defect LS theory in IR.

\begin{figure}[h]
\centering
\includegraphics[scale=0.65]{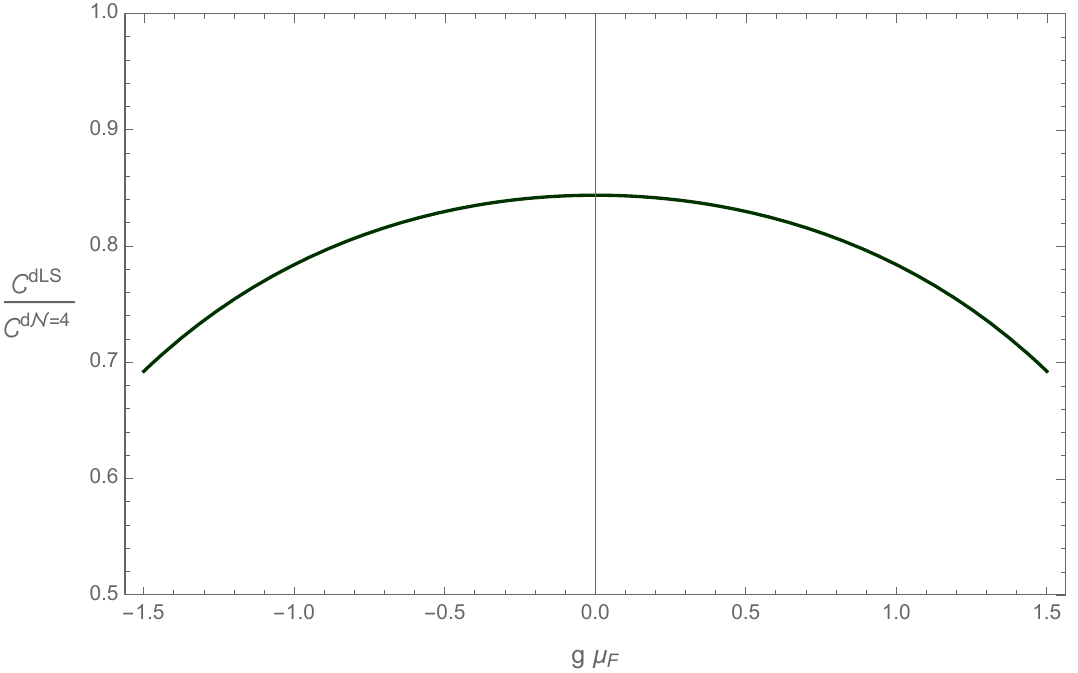}
\caption{The ratio of $\mathcal{C}^{(4)}$ between LS and $\mathcal{N}=4$ (with $\mu_B=0$)  defects for $n=1$. This quantity is less than one for all allowed values of the monodromy sources.}\label{fig:Cd5Dratnis1}
\end{figure}

\begin{figure}[h]
\centering
\includegraphics[scale=0.75]{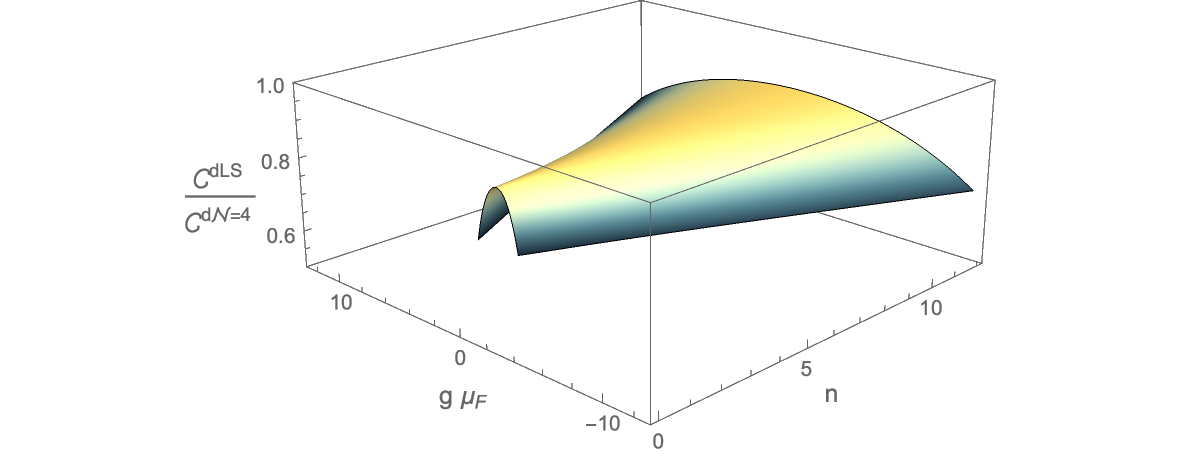}
\caption{The ratio of $\mathcal{C}^{(4)}$ between LS and $\mathcal{N}=4$ (with $\mu_B=0$)  defects for various values of $n\ge1$. This quantity is less than one for all allowed values of the monodromy sources.}\label{fig:Cd5Dratnis1}
\end{figure}

\end{appendix}

\bibliography{defects}{}
\bibliographystyle{utphys}

\end{document}